\documentclass[fleqn,usenatbib]{mnras}

\usepackage{newtxtext,newtxmath}

\usepackage[T1]{fontenc}
\usepackage{ae,aecompl}
\usepackage{tikz}
\usepackage{bm}
\usepackage{relsize}
\usetikzlibrary{positioning}
\usepackage{listings}
\usepackage{xcolor} 

\usepackage{graphicx}	
\usepackage{amsmath}	
\usepackage{amssymb}	
\usepackage{color}
\definecolor{linkcolor}{rgb}{0,0,0.25}
\hypersetup{
  colorlinks=true,        
  linkcolor=linkcolor,    
  citecolor=linkcolor,    
  filecolor=linkcolor,    
  urlcolor=linkcolor      
}

\makeatletter
\renewcommand{\@printed}{}
\makeatother

\newcommand{\figurename}{Figure}
\newcommand{\tablename}{Table}
\newcommand{\eqnname}{Equation}

\definecolor{darkgreen}{rgb}{0.0, 0.5, 0.0}

\definecolor{darkblue}{rgb}{0.0, 0., 0.7}

\newcommand{\teff}{\ensuremath{T_\mathrm{eff}}}
\newcommand{\logg}{\ensuremath{\log g}}
\newcommand{\xh}[1]{\ensuremath{[\mathrm{#1/H}]}}
\newcommand{\xfe}[1]{\ensuremath{[\mathrm{#1/Fe}]}}
\newcommand{\madstd}{\ensuremath{\sigma^{\mathrm{MAD}}}}
\newcommand{\plx}{\ensuremath{\varpi}}
\newcommand{\mas}{\ensuremath{\mathrm{mas}}}
\newcommand{\uas}{\ensuremath{\mu\mathrm{as}}}

\newcommand{\gaia}{\emph{Gaia}}

\title[Spectro-photometric distances and the \gaia\ DR2 zero-point]{Simultaneous calibration of spectro-photometric distances and the \gaia\ DR2 parallax zero-point offset with deep learning}

\author[Leung \& Bovy]{
Henry W. Leung$^{1}$\thanks{E-mail: henrysky.leung@mail.utoronto.ca}
\& Jo Bovy$^{1,2}$\thanks{Alfred P. Sloan Fellow}
\\
$^{1}$Department of Astronomy and Astrophysics, University of Toronto, 50 St. George Street, Toronto, Ontario, M5S 3H4, Canada\\
$^{2}$Dunlap Institute for Astronomy and Astrophysics, University of Toronto, 50 St. George Street, Toronto, Ontario, M5S 3H4, Canada
}

\date{}

\pubyear{2019}

\begin{document}
\label{firstpage}
\pagerange{\pageref{firstpage}--\pageref{lastpage}}
\maketitle

\begin{abstract}
\gaia\ measures the five astrometric parameters for stars in the Milky Way, but only four of them (positions and proper motion, but not parallax) are well measured beyond a few kpc from the Sun. Modern spectroscopic surveys such as APOGEE cover a large area of the Milky Way disk and we can use the relation between spectra and luminosity to determine distances to stars beyond \gaia's parallax reach. Here, we design a deep neural network trained on stars in common between \gaia\ and APOGEE that determines spectro-photometric distances to APOGEE stars, while including a flexible model to calibrate parallax zero-point biases in \gaia\ DR2. We determine the zero-point offset to be $-52.3 \pm 2.0\,\uas$ when modeling it as a global constant, but also train a multivariate zero-point offset model that depends on $G$, $G_{BP}-G_{RP}$ color, and \teff\ and that can be applied to all $\approx 139$ million stars in \gaia\ DR2 within APOGEE's color--magnitude range. Our spectro-photometric distances are more precise than \gaia\ at distances $\gtrsim 2\,\mathrm{kpc}$ from the Sun. We release a catalog of spectro-photometric distances for the entire APOGEE DR14 data set which covers Galactocentric radii $2\,\mathrm{kpc} \lesssim R \lesssim 19\,\mathrm{kpc}$; $\approx 150,000$ stars have $<10\%$ uncertainty, making this a powerful sample to study the chemo-dynamical structure of the disk. We use this sample to map the mean \xh{Fe} and 15 abundance ratios \xfe{X} from the Galactic center to the edge of the disk. Among many interesting trends, we find that the bulge and bar region at $R \lesssim 5\,\mathrm{kpc}$ clearly stands out in \xh{Fe} and most abundance ratios.
\end{abstract}





\section{Introduction}

The Milky Way provides a unique opportunity for the study of galaxy formation and evolution, because we can determine the three-dimensional position and velocity (using astrometry and spectroscopy), high-quality stellar parameters and elemental abundances (from high-resolution spectroscopy), and ages for large samples of individual stars \citep[e.g.,][]{Freeman02a,Rix13a}. Recently, great advances have been made in getting precise stellar parameters \citep[e.g.,][]{Holtzman15a,2016arXiv160303040C,2019MNRAS.483.3255L} and ages \citep[e.g.,][]{Martig16a,Ness16a,Mackereth19a} for hundreds of thousands of stars across the Milky Way from high-resolution spectroscopic surveys such as APOGEE \citep{Majewski17a}. At the same time, astrometric data from the \gaia\ satellite \citep{2016A&A...595A...1G} are providing an unprecedented view of the spatial and kinematic structure of the extended solar neighborhood \citep[e.g.,][]{2018A&A...616A..11G,2018Natur.561..360A,2019MNRAS.482.1417B}. However, even with \gaia's exquisite astrometric precision, it currently only provides precise distances (and, thus, tangential motions) within about 2 kpc, even for relatively bright giants. Thus, to take full advantage of the wide disk coverage of APOGEE, we require a method for obtaining precise distances and space velocities for \emph{all} stars in APOGEE. The wealth of data from \gaia\ allows spectro-photometric distance methods to be calibrated using the large, nearby set of \gaia\ parallaxes and then be applied to the full APOGEE data set. This is what we set out to do in this paper, using the modern machine-learning technique \emph{deep learning}.

Machine-learning techniques for spectro-photometric distances like deep learning are powerful, because they can be trained on stars with both high-resolution spectra from APOGEE and parallaxes from \gaia\ to produce spectro-photometric distances. But applied blindly, such techniques propagate biases that are present in the training set to the model and the subsequently inferred distances. In particular, the \gaia\ DR2 parallaxes are known to suffer from a zero-point offset \citep{2018A&A...616A...2L,2018arXiv180502650Z} that is known to be multivariate and may have a very complex dependence on magnitude, color, sky position, or other quantities. Simply training on the parallax data without any correction or using an inappropriate correction will result in a biased model. This would significantly bias distances obtained for distant stars such as those in the APOGEE sample, because these have small parallaxes where even a small (tens of \uas) zero-point offset has a large effect.

Determinations of the \gaia\ DR2 parallax zero-point offset have either been performed using quasars \citep{2018A&A...616A...2L}, which should have no measurable parallax on average, or various types of stars. The determination using quasars is precise, but likely not directly applicable to most stars of interest, which are both brighter and bluer than quasars and thus probably have a different zero-point offset due to its multivariate dependencies. Determinations using stars depend on trusting semi-empirical stellar-evolution models \citep{2018arXiv180502650Z} or use rigid (i.e., few parameter) models for standardizable candles \citep[e.g.,][]{2017ApJ...838..107S,2018ApJ...861..126R}. These give precise determinations of the zero-point, but have to assume that the zero-point is a well-behaved (e.g., constant) function of observables such as magnitude, parallax, and color. In this paper, we determine the parallax zero-point offset using a diverse sample of main-sequence and red-giant stars with a method that only rests on the assumption that continuum-normalized stellar spectra allow the intrinsic luminosity to be determined, but not the distance (or apparent magnitude). We represent the zero-point's multivariate dependence on magnitude, color, and temperature using a flexible (i.e., many parameter) neural network model using deep learning. 

Deep learning is a subset of machine learning and the term refers to the usage of multi-layer (``deep'') artificial neural networks (NN) to do various kinds of machine learning tasks in supervised and unsupervised learning, image recognition, natural language processing, etc. NNs exist in various architectures that mimic biological brains in order to represent high-dimensional mappings in a versatile manner. On top of the versatility of the NN, we also employ in our work a robust way of Bayesian deep learning that (a) takes data uncertainties in the training data into account and (b) estimates uncertainty in predictions made with the NN using an approximation to variational inference, drop out variational inference. This methodology is mostly based on our previous work on deep learning of stellar abundances with APOGEE spectra using \texttt{astroNN}\footnote{\url{https://github.com/henrysky/astroNN}} \citep{2019MNRAS.483.3255L}. Supervised learning requires trusted, labeled training data for the model to learn. Because of the \gaia\ DR2 zero-point offset, we do not fully trust the training data in the current application and in training a NN to determine spectro-photometric distances from APOGEE spectra, we need to simultaneously learn how to correct the \gaia\ parallaxes for the zero-point offset. We do this by using a form of adverserial training that optimizes \emph{two} NNs at the same time, one for the spectro-photometric distances and one for the zero-point calibration, in such a way that the residual between the spectro-photometric parallax and the calibrated \gaia\ parallax has no information about the parallax itself.

The outline of this paper is as follows. Section \ref{sec:methodology} describes our methodology, with Section \ref{subsec:assumption} focusing on a general discussion of the method and assumptions and Section \ref{subsec:models} discussing the specifics of the model and its implementation in more detail. Section \ref{sec:data} provides information on the data selection and processing from APOGEE DR14 and \gaia\ DR2 to construct training and test sets. Section \ref{sec:offset_result} and Section \ref{sec:precision} discuss our \gaia\ DR2 zero-point offset findings and the precision of the spectro-photometric distances, respectively. To illustrate the power of the derived data set of APOGEE stars with precise distances, Section \ref{sec:MW_chem} shows maps of the elemental abundances across a wide area of the Milky Way. Section \ref{sec:discuess} discusses how our method compares to other methods for inferring the parallax zero-point and for determining spectro-photometric distances, and we look forward to future applications of this methodology. Section \ref{sec:conclusion} gives a brief overview of our conclusions. Appendix \ref{append:A} describes how to perform variational NN inference on arbitrary APOGEE spectra to determine their distances using the model used in this work.

Code to reproduce all of the plots in this paper as well as the \texttt{FITS}\footnote{\url{https://github.com/henrysky/astroNN_gaia_dr2_paper/raw/master/apogee_dr14_nn_dist.fits}} data file containing our neural network's distance for the entire APOGEE DR14 data set is available at \url{https://github.com/henrysky/astroNN_gaia_dr2_paper}. The data model for this \texttt{FITS} file is decribed in \tablename~\ref{table:data_model} at the end of this paper.

\section{Methodology} \label{sec:methodology}

\subsection{Basic idea and assumptions}\label{subsec:assumption}

Our goal is to simultaneously calibrate spectro-photometric distances and the \gaia\ DR2 parallax zero-point offset by training a model to predict the \gaia\ DR2 catalog value of the parallax from the near-infrared APOGEE spectra. To do this, our supervised deep-learning algorithm requires training labels for a set of stars observed by both \gaia\ and APOGEE and training proceeds by minimizing the prediction error for the training data set by adjusting the model parameters. In the present application, these model parameters are the strengths of neural-network connections and the optimization is performed using a gradient-descent optimizer. 

Before explaining our method in detail, we briefly summarize the main ideas and assumptions that allow us to calibrate both the spectro-photometric distance model and the \gaia\ DR2 zero-point offset at the same time:\\
\emph{Spectro-photometric luminosity features:} Our spectro-photometric distance model works by mapping continuum-normalized spectra to luminosity (or absolute magnitude) that is subsequently converted to parallax using the observed apparent magnitude and the extinction. To perform this mapping from spectra to luminosity without relying on stellar evolution models, we have to assume that the continuum-normalized spectrum contains features that are indicative of the star's luminosity. This is a plausible assumption for the red giants observed by APOGEE, as the mass-dependence of internal mixing of carbon and nitrogen in red giants allows the stellar mass to be determined from APOGEE spectra \citep[e.g.,][]{Masseron15a} and mass combined with the effective temperature and surface gravity that are eminently measurable from stellar spectra allow the luminosity to be determined. It is less clear that this is plausible for the sub-giant, turn-off, and main-sequence stars in our sample, but ultimately the success (or lack thereof) of the method validates this assumption and we will see that we are able to determine luminosities for these types of stars as well.\\
\emph{Continuous spectral flux--luminosity relation:} We assume that the value of the luminosity is a continuous, smooth function of the continuum-normalized spectral flux values. This is a general limitation of the type of NN that we use. Thus, similar spectra are assumed to have similar luminosities. This is a generalization of the concept that ``spectral twins'' have the same intrinsic luminosity, which has been successfully used to obtain high-precision distances to stars \citep{Jofre15a,Jofre17a}. However, we do not require that spectra with similar luminosities have similar spectra.\\
\emph{Unpredictability of the distance from continuum-normalized spectra:} This is the basic assumption that allows the calibration of the \gaia\ DR2 zero-point offset. We assume that a star's distance cannot be learned from the continuum-normalized spectra alone. If this assumption did not hold, then the simultaneous determination of the luminosity and distance from the continuum-normalized spectrum would allow the parallax and apparent magnitude be determined directly and the spectro-photometric model on its own could match the \gaia\ catalog parallax values at all magnitudes and colors, thus absorbing the erroneous effect of the zero-point offset. That the continuum-normalized spectrum contains no direct distance information is plausible, because the in the absence of interstellar absorption or emission features, the continuum-normalized spectrum is an intrinsic, distance-independent property of the star. APOGEE spectra do contain interstellar absorption features (e.g., a strong diffuse interstellar band; \citealt{2015ApJ...798...35Z}), which are correlated with extinction and thus indirectly with distance and this therefore weakly breaks our assumption. However, as long as the correlation with distance is not perfect, the effect of this on our method is small. We similarly assume that the continuum-normalized spectra have no features that are directly related with the zero-point offset. That a spectrum would have features directly related to the zero-point offset is highly unlikely, given that the zero-point offset is a \gaia-specific instrumental effect and the spectra are taken using an entirely different instrument. The unpredictability of the distance from continuum-normalized spectra leads therefore to the unpredictability of the \gaia\ zero-point offset from spectra.\\
\emph{Gaussian uncertainty of Gaia parallaxes:} We assume that the uncertainty value provided by the \gaia\ data reduction pipeline is the standard deviation of a Gaussian distribution of parallax error. Below, we preserve the Gaussian nature of the uncertainty by not performing any non-linear (e.g., inverse or logarithmic) operation on the parallax, but instead adopting the luminosity--parallax relation from \cite{2018AJ....156..145A}
\begin{equation} \label{eq:fakemag}
L_\mathrm{fakemag} = \plx 10^{\frac{1}{5}m_\mathrm{apparent}} = 10^{\frac{1}{5}M_\mathrm{absolute}+2}\,,
\end{equation}
where $L_\mathrm{fakemag}$ is an alternative scaling of luminosity, a pseudo-luminosity. If the uncertainty in the parallax is Gaussian and the uncertainty in apparent magnitude is negligible, the uncertainty in $L_\mathrm{fakemag}$ is Gaussian as well.\\
\emph{Known extinction:} The conversion between intrinsic luminosity and parallax requires the extinction-free apparent magnitude. In this work, we assume that the extinction is known and that there is no uncertainty in the extinction correction. This is a good assumption for APOGEE stars, for which the $K_{s}$ extinctions are determined from near-infrared and mid-infrared photometry (see below). Note that this is not a crucial assumption of the method, because we could simultaneously infer extinction from the spectra and associated multi-band photometric data \citep[e.g.,][]{2018arXiv181009468H}, but because accurate extinctions are available for APOGEE stars (see below), it is a convenient additional assumption for the present application.

We exploit the unpredictability of the \gaia\ zero-point offset from continuum-normalized spectra to train a model that we name the \emph{offset calibration model}. This model calibrates the zero-point offset during training while another part of the model, the \emph{spectro-luminosity model} learns how to map continuum-normalized spectra to luminosity. Since the goal of the training is to minimize the prediction error for the training data---in our case obtaining a predicted \gaia\ parallax as close to the actual \gaia\ parallax as possible given the error in the observed parallax---any offset that is unpredictable to the spectro-luminosity model will result a higher prediction error. The offset calibration model then effectively provides one or more degrees of freedom in parallax space to improve the parallax prediction. Given our assumptions, the spectro-luminosity model returns the true parallax and the offset calibration model therefore provides the zero-point offset.

\subsection{Modeling specifics}\label{subsec:models}

\begin{figure*}
\centering
\includegraphics[width=\textwidth]{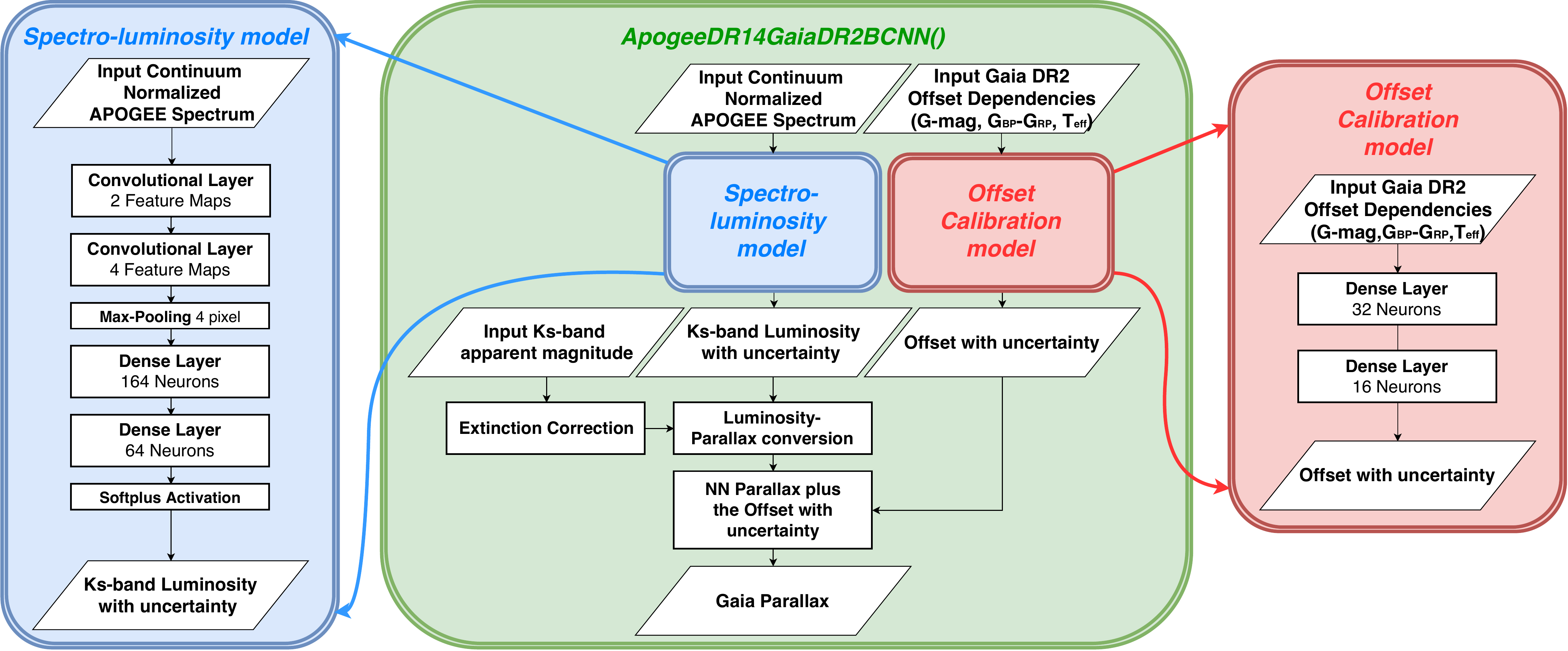}
\caption{The neural-network architecture used in this work. The full model is defined as \texttt{ApogeeDR14GaiaDR2BCNN()} in \texttt{astroNN} (middle panel) and it combines two separate NNs: a \emph{spectro-luminosity model} to map APOGEE spectra to luminosity (right panel) and an \emph{offset calibration model} to predict the \gaia\ DR2 parallax zero-point offset based on the stellar properties $G$, $G_{BP}-G_{RP}$, and \teff. The \texttt{Luminosity-Parallax Conversion} layer refers to the use of \eqnname~\eqref{eq:fakemag} for our pseudo-luminosity definition. The implementation of this model in \texttt{astroNN} behaves differently during training and testing. During training, all of the layers are used, i.e. \texttt{ApogeeDR14GaiaDR2BCNN()} is trained as a whole to predict the \gaia\ catalog value of the parallax. During testing or when applying the model to new data, the \emph{spectro-luminosity model} is used separately when determining luminosity and distance for a given APOGEE spectrum. Similarly, the offset calibration model can be used separately to give the zero-point offset and correct the \gaia\ parallax for any star in \gaia\ DR2 with similar magnitudes and colours to those in APOGEE.}
\label{figure:nn_flow}
\centering
\end{figure*}

The NN that we train and test in this work is composed of two basic components: the \emph{offset calibration model} and the \emph{spectro-luminosity model}. Both of these are neural networks and they are trained simultaneously using a single objective function. Once these two models are trained they are used separately to either determine the luminosity (and resulting distance) from an APOGEE spectrum or to predict the \gaia\ zero-point offset for a given star observed by \gaia.

The \emph{spectro-luminosity model}---\texttt{ApogeeBCNN()} in \texttt{astroNN}---maps continuum-normalized APOGEE spectra to the pseudo-luminosity of Equation~\eqref{eq:fakemag} in the 2MASS $K_{s}$ band. When combined with the extinction-corrected apparent $K_{s}$ magnitude the distance or parallax follows. The offset calibration model provides the \gaia\ DR2 zero-point offset. In the most general form that we use below, the offset calibration model maps observed quantities such as the \gaia\ $G$ band magnitude, $G_{BP}-G_{RP}$ colour, and effective temperature to a zero-point offset. In this case, \emph{each individual star gets its own zero-point offset}. We also consider the case where the zero-point is a simple constant that is fit in the training. We implement this model using the same NN model as in the multivariate case, except that the inputs are set to the same value for each star. This therefore produces the same zero-point for all stars, but implementing it this way allows us to determine the uncertainty in the constant zero-point offset using dropout variational inference. The spectro-luminosity also returns a predictive variance for each star individually that represents the uncertainty in the model fit (see below).

The architecture of the \emph{spectro-luminosity model} is shown in the left-hand panel of \figurename~\ref{figure:nn_flow}; it consists of a sequence of convolutional layers, a max-pooling layer, a few dense layers, and a \texttt{Softplus} output activation function (discussed further below). The architecture of the offset calibration model is displayed in the right panel of \figurename~\ref{figure:nn_flow} and it consists only of dense layers.

To train these two NNs simultaneously, we optimize the NNs to correctly predict the \gaia\ DR2 catalog value of the parallax. That is, we aim to predict the parallax as \gaia\ reports it, including the zero-point offset. The full model architecture that combines the spectro-luminosity model and the offset calibration model is defined as \texttt{ApogeeDR14GaiaDR2BCNN()} in \texttt{astroNN} and is shown in \figurename~\ref{figure:nn_flow}. This combined network takes the continuum-normalized APOGEE spectrum for a star $i$ and maps it to the $\hat{L}_\mathrm{fakemag}$ pseudo-luminosity\footnote{In this section we denote predicted values of quantities using the hat operator (e.g., $\hat{\plx}$ for the predicted parallax) to distinguish them from the provided training data. In later sections we drop this notation and simply denote predicted parallaxes and parallax offsets without a hat.}, which is converted to the true parallax $\hat{\plx}_i$ using the observed apparent magnitude and the known extinction. The offset calibration model provides the zero-point offset $\hat{\plx}_{\mathrm{offset},i}$ (either for each star individually or simply as a constant). This offset is added to the true parallax to provide the predicted \gaia\ DR2 parallax
\begin{equation}
    \hat{\plx}_i^G = \hat{\plx}_i + \hat{\plx}_{\mathrm{offset},i}\,.
\end{equation}

The objective function that is minimized during training is then the sum of the following objective functions for each individual star $i$
\begin{equation} \label{eq:mmse}
   J(\plx_i^G, \hat{\plx}_i^G) = \frac{1}{2} (\hat{\plx}_i^G-\plx_i^G)^2 e^{-s_i} + \frac{1}{2}\,s_i
\end{equation}
where $\plx_i^G$ is the \gaia-reported parallax. This objective function is minus the log likelihood if the uncertainty in the difference is Gaussian, which we assume here. The quantity $s_i$ represents the natural logarithm of the uncertainty variance in the difference between the predicted and the \gaia\ parallax. It has two contributions that are summed in quadrature
\begin{equation}\label{eq:total_var}
s_i =  \ln \left[\sigma^2_{\plx,i} + \sigma^2_{\mathrm{pred}\,\plx,i}\right]\,.
\end{equation}
These two contributions are the parallax uncertainty $\sigma^2_{\plx,i}$ reported by \gaia\ and the predictive uncertainty $\sigma^2_{\mathrm{pred}\,\plx,i}$ in the true parallax returned by the spectro-luminosity model. This predictive uncertainty is an additional output from the spectro-luminosity model and the optimal mapping of input continuum-normalized spectrum to uncertainty is also simultaneously optimized during training. The spectro-luminosity model returns $L_\mathrm{fakemag}$ and its (Gaussian) uncertainty, but this is easily converted to $\sigma^2_{\mathrm{pred}\,\plx,i}$ using \eqnname~\eqref{eq:fakemag}. The purpose of including the predictive variance is to capture the uncertainty associated with the inability of the model to perfectly map spectra onto absolute magnitudes. We do not include a predictive uncertainty in the offset calibration model, because for predictions we are primarily interested in spectro-photometric distances. The final loss for the stochastic gradient descent optimizer in this work (ADAM optimizer; \citealt{2014arXiv1412.6980K}) is calculated from a mini-batch partition of the data of size $N$
\begin{equation}
    J = \frac{1}{N}\sum_{i=1}^N J(\plx_i^G, \hat{\plx}_i^G)\,.
\end{equation}
We use dropout with a dropout fraction of 30\% in all layers during training to prevent overfitting \citep{2012arXiv1207.0580H}.

\begin{figure*}
\centering
\includegraphics[width=\textwidth]{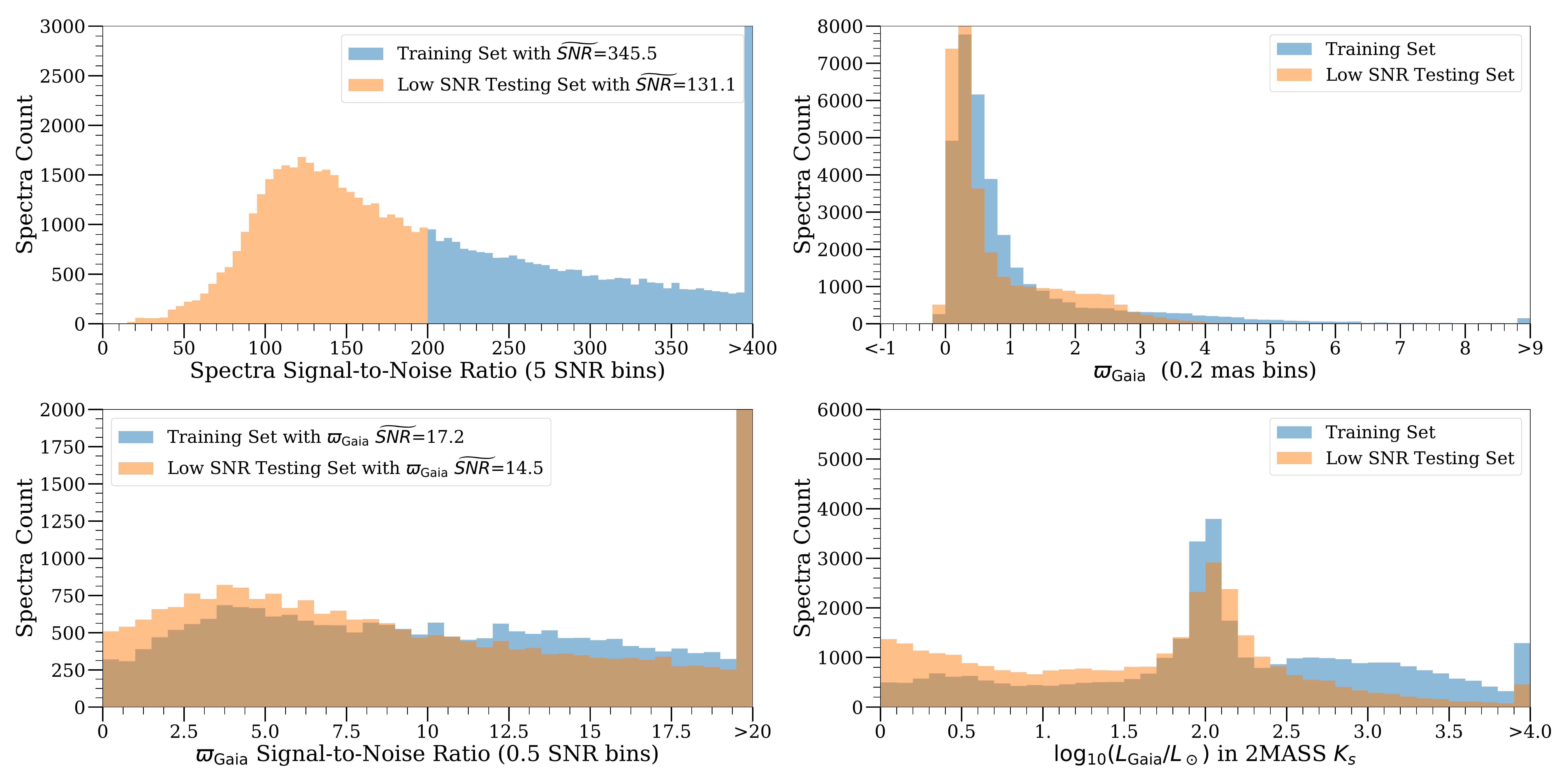}
\caption{Number of stars in the training and test sets as a function of four different properties. Upper left: APOGEE spectral SNR with the median SNR $\widetilde{SNR}$ shown in the legend. Upper right: observed \gaia\ parallax, without any zero-point offset correction. Bottom left: \gaia\ parallax SNR (we do not cut on parallax SNR to construct our training and test sets). Bottom right: logarithmic luminosity in solar units, derived from unmodified \gaia\ parallax and extinction-corrected $K_s$ apparent magnitude; we exclude stars below solar luminosity.}
\label{figure:SNR}
\centering
\end{figure*}

During inference, we use dropout for uncertainty estimation which is known as dropout variational inference \citep{2017arXiv170304977K}. To obtain $L_\mathrm{fakemag}$ predictions for input spectra, we run every spectrum through $N$ forward passes of the spectro-luminosity model with dropout turned on. Since dropout drops weights randomly, the spectro-luminosity model becomes probabilistic and has different predictions in every forward pass through the network. The mean value of these $N$ predictions is the final prediction and the standard deviation of the predictions is the model uncertainty. In addition to this model uncertainty from dropout variation inference, the spectro-luminosity model also gives the predictive uncertainty for each star discussed above. The total prediction uncertainty is the sum of model and predictive uncertainty in quadrature \citep{2017arXiv170304977K}.

The reason that we use the \texttt{Softplus} activation in the last layer of the spectro-luminosity model is that the pseudo-luminosity $L_\mathrm{fakemag}$ from \eqnname~\eqref{eq:fakemag} cannot be negative, because a negative luminosity would translate to a negative true parallax. The \texttt{Softplus} activation---defined as $y = \log{(1 + e^x)}$ where x is the input and y is the output---is a smooth approximation of the standard rectified linear unit (\texttt{ReLU}) activation function that we use in all but the output layer. Although both \texttt{Softplus} and \texttt{ReLU} map all real inputs to non-negative real outputs, using the \texttt{Softplus} activation as the last layer for predicting stellar luminosity is better, because it never produces zero luminosity and thus never leads to zero parallax. That is, unlike \texttt{ReLU}, \texttt{Softplus} maps all real numbers to non-zero positive real numbers, while \texttt{ReLU} maps all negative numbers to zero. 

While the true parallax returned by our model is never negative or zero, our model's prediction for the \gaia\ parallax can be negative due to the zero-point correction. This in addition to taking the uncertainty in the parallax and its prediction into account in the objective function of \eqnname~\eqref{eq:mmse} allows us to use negative \gaia\ parallaxes in our training set. That is, we do not need to artificially remove negative parallaxes in the \gaia\ catalog, which may result from random noise or from zero-point biases. Because our objective function takes the \gaia\ parallax uncertainty into account, we are also not limited to only using high precision \gaia\ parallaxes in the training and we do not do any cut on parallax signal-to-noise ratio in the training step.

\section{Data}\label{sec:data}

\subsection{Spectroscopic data from APOGEE}\label{subsec:apogee}

The spectroscopic data in this work come from Data Release 14 (DR14; \citealt{2018arXiv180709773H,2018arXiv180709784J,2018ApJS..235...42A}) of the APO Galactic Evolution Experiment (APOGEE;   \citealt{Majewski17a}). APOGEE spectra are obtained with a 300-fiber spectrograph \citep{2019arXiv190200928W} attached to the Sloan Foundation 2.5m telescope at Apache Point Observatory \citep{2006AJ....131.2332G}. APOGEE is an infrared ($1.5 \mu m$ to $1.7 \mu m$), high resolution ($R\sim 22,500$), high signal-to-noise ratio (typical $\mathrm{SNR}>100$) spectroscopic survey. As in our previous work \citep{2019MNRAS.483.3255L}, we perform our own continuum-normalization that uses a pre-defined set of continuum pixels \citep{2016arXiv160303040C,2016ApJ...817...49B} starting from the combined, rest-frame spectra in the APOGEE ``apStar'' files. After continuum normalization, we set the flux value of pixels that contain the following bits in the APOGEE
pixel-level mask bits in the \texttt{APOGEE\_PIXMASK} bitmask to 1 (the expected continuum), because they are likely bad: {\bf 0}: bad pixel, {\bf 1}: cosmic ray, {\bf 2}: saturated, {\bf 3}: unfixable, {\bf 4}: bad from dark, {\bf 5}: bad from flat, {\bf 6}: high error, {\bf 7}: no sky info, {\bf 12}: overlaps a significant sky line. Spectroscopic parameters such as \teff\ and elemental abundances that we use come from our own re-analysis of the DR14 data in \citet{2019MNRAS.483.3255L}.

Apparent $K_{s}$ magnitudes for all stars in the APOGEE catalog are taken from the 2MASS catalog \citep{2006AJ....131.1163S}. We correct these for extinction using the \texttt{AK\_TARG} extinction listed in the APOGEE catalog, which is the extinction adopted by APOGEE for targeting. The value of \texttt{AK\_TARG} is derived on an individual-star basis by the Rayleigh Jeans Color Excess method \citep[RJCE; ][]{2011ApJ...739...25M}. The RJCE method calculates extinctions using a combination of near- and mid-infrared photometry as 
\begin{equation} \label{eq:rjce}
A_{Ks} = 0.918 (H - [4.5\mu] - (H - [4.5\mu])_0)
\end{equation}
where $H - [4.5\mu]$ is the measured color and the method assumes that $(H - [4.5\mu])_0=0.08$ for a wide range of spectral types. $H$-band photometry in this equation comes from 2MASS, while the $4.5\mu m$ photometric data are either from Spitzer-IRAC data \citep{2009PASP..121..213C} or from the WISE survey \citep{2010AJ....140.1868W}. We set all \texttt{AK\_TARG} $<-1$ equal to 0, that is, zero extinction.

\subsection{Parallax data from \gaia}

We use data from the second data release (DR2; \citealt{GaiaDR2}) from the European Space Agency's Gaia mission \gaia\ \citep{2016A&A...595A...1G} to train the model and calibrate the \gaia\ DR2 zero-point offset. \gaia\ DR2 contains 1.3 billions source with 5 astrometric parameters (positions, proper motions, parallaxes). We match \gaia\ sources to APOGEE stars using a celestial position cross-match with a tolerance of $2^{\prime\prime}$. Out of 277,371 stars in APOGEE DR14, 265,761 have both a \gaia\ DR2 parallax and a 2MASS $K_{s}$ band apparent magnitude; the median parallax signal-to-noise ratio of these stars is 15.4. We use parallax and \gaia\ pipeline-reported parallax uncertainty as well as the $G$-band magnitude and $G_{BP}-G_{RP}$ color.

\subsection{Training, validation, and test data sets from APOGEE and \gaia}\label{subsec:sets}

We create one training set and one test set from the APOGEE DR14 spectra. Both data sets consist of continuum-normalized APOGEE spectra and \gaia\ DR2 parallaxes. The training set contains 35,112 stars, while the test set has 33,468 stars. The main difference between the training and test set is in the signal-to-noise (SNR). The training is constructed using only high-SNR spectra with SNR>200, whereas the test set consists only of low-SNR spectra with SNR<200. $90\%$ of the training set is randomly selected to train the NN--that is, these stars are used to compute the gradients of the objective function in the training steps---and the remaining $10\%$ constitutes a separate validation set that is used to validate the performance of the NN during the training process.

On top of the spectral SNR cuts, we perform cuts on the quality of the \gaia\ parallaxes and on the quality of the APOGEE spectra. This is necessary, because all of the knowledge learned by the NN is solely driven by the training data. Therefore, we need to make sure that the training inputs and labels are as accurate as possible, because any systematic inaccuracy such as bias will be captured by the NN and propagated to new data. For this reason, we exclude spectra flagged with the \texttt{STARFLAG} flags and spectra with a radial velocity scatter larger than $1\,\mathrm{km\,s}^{-1}$, because these represent potential issues with the spectra, and potential binary stars, respectively. We require that the \gaia\ parallaxes have $\sigma_{\plx}$< 0.1\,mas and \texttt{visibility\_periods\_used} $\geq11$  to ensure stars are astrometrically well-observed by \gaia\ with at least 11 gaps of at least 4 days with small uncertainty \citep{2018A&A...616A...2L} . Furthermore, we exclude stars with  $\log_{10}{\left(L^{K_s}_{\mathrm{Gaia}}/L_\odot\right)}\leq0$ where $L$ refers to the use of $L_\mathrm{fakemag}$ in \eqnname~\ref{eq:fakemag}, because we are mainly interested in brighter stars that can be seen to large distances to map the Milky Way. These cuts ensure the quality of the training and test sets. \figurename~\ref{figure:SNR} displays the distribution of the training and test sets in a few key quantities.

Despite all of our quality cuts, our training and test sets include negative parallaxes and parallaxes with large percentage uncertainties (low parallax SNR). As described in Section \ref{subsec:models}, our model can handle negative parallaxes in a physically sensible way during the training process and as discussed in Section \ref{subsec:assumption}, our training process does not involve inverse parallax, which is a strongly biased estimate of the distance for low parallax SNR \citep{2015PASP..127..994B}. Furthermore, our robust objective function used during training takes parallax uncertainty into account. Therefore, we do not remove parallaxes with low SNR from the training or test set and are therefore not affected by any biases that would result from making such a cut.

\section{Results: The \gaia\ DR2 zero-point offset}\label{sec:offset_result}

\begin{figure*}
\centering
\includegraphics[width=0.8\textwidth]{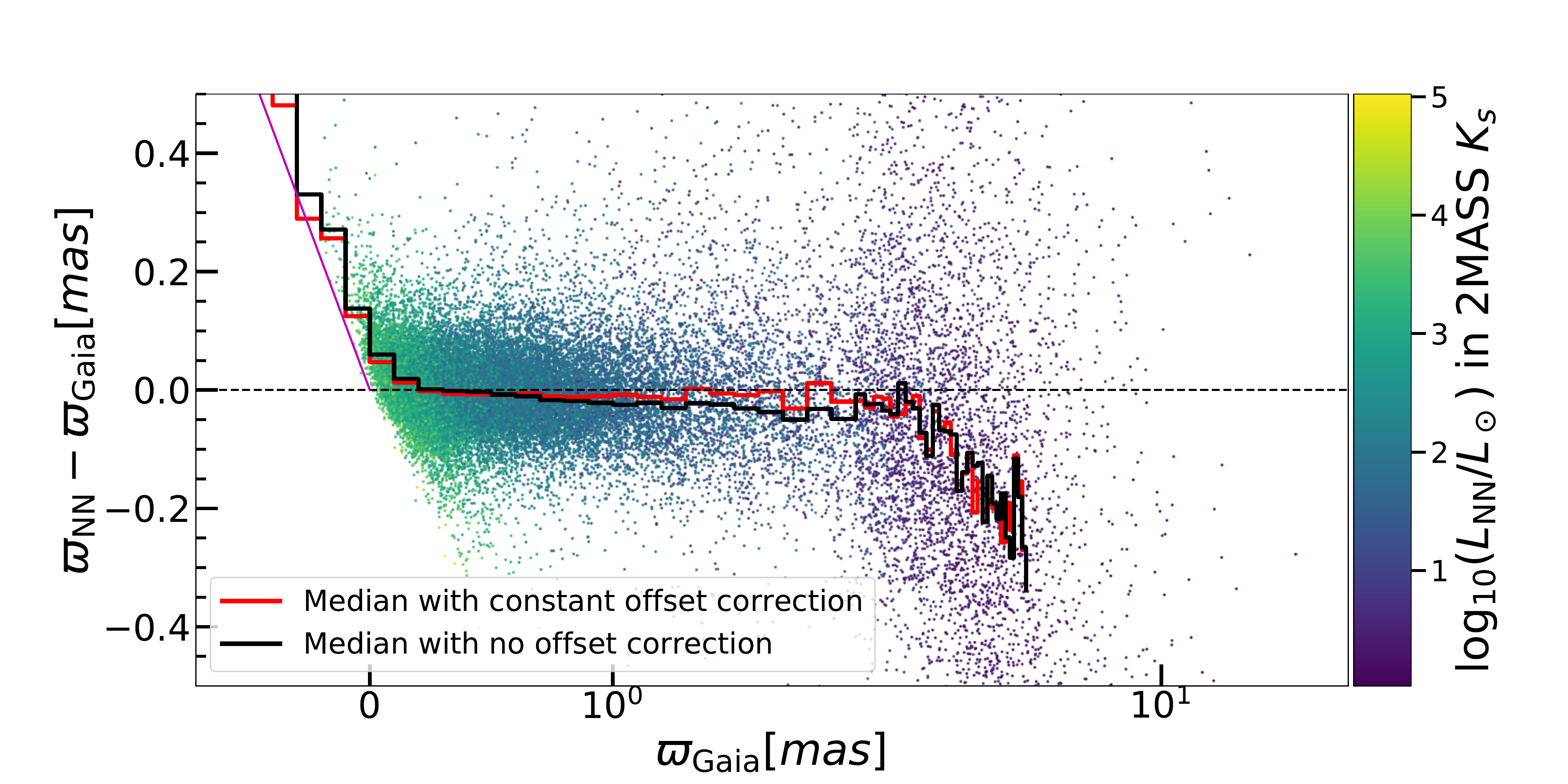}
\caption{Comparison of the neural network spectro-photometric parallax results obtained by training with unmodified \gaia\ parallaxes and to those obtained using the constant offset calibration model for the test set. The $x$ axis is the \gaia\ DR2 catalog value for the parallax and the $y$ axis is the difference between the spectro-photometric parallax and the \gaia\ DR2 parallax, either the catalog value for the model without zero-point correction or the zero-point-corrected value $\plx_{\mathrm{Gaia}}+52\,\uas$ for the model trained with a constant offset calibration model. The scatter points are the prediction from the constant offset calibration model. The curves show the median of the parallax difference values in bins in $\plx_\mathrm{Gaia}$. The purple line starting at $\plx_\mathrm{Gaia}=0$ is the line $\plx_\mathrm{NN} = 0$, which is the smallest the NN parallax can be and therefore no points can lie below this line. That the curves follow this line is because the NN returns a parallax close to zero for most stars for which \gaia\ determined a negative parallax. Within a few 100 pc from the Sun ($\plx \gtrsim 3\,\mathrm{mas}$)  the APOGEE stars are mainly dwarfs and the model overpredicts their luminosity, because the training set is dominated by giants; this creates the downward trend at high parallax. Overall, the model with constant zero-point offset is better able to match the \gaia\ parallaxes over a wide range of parallax values than the model without a zero-point correction, demonstrating that a zero-point offset is present in the \gaia\ DR2 parallaxes.}
\label{figure:result_no_correct}
\centering
\end{figure*}

\begin{figure*}
\centering
\includegraphics[width=\textwidth]{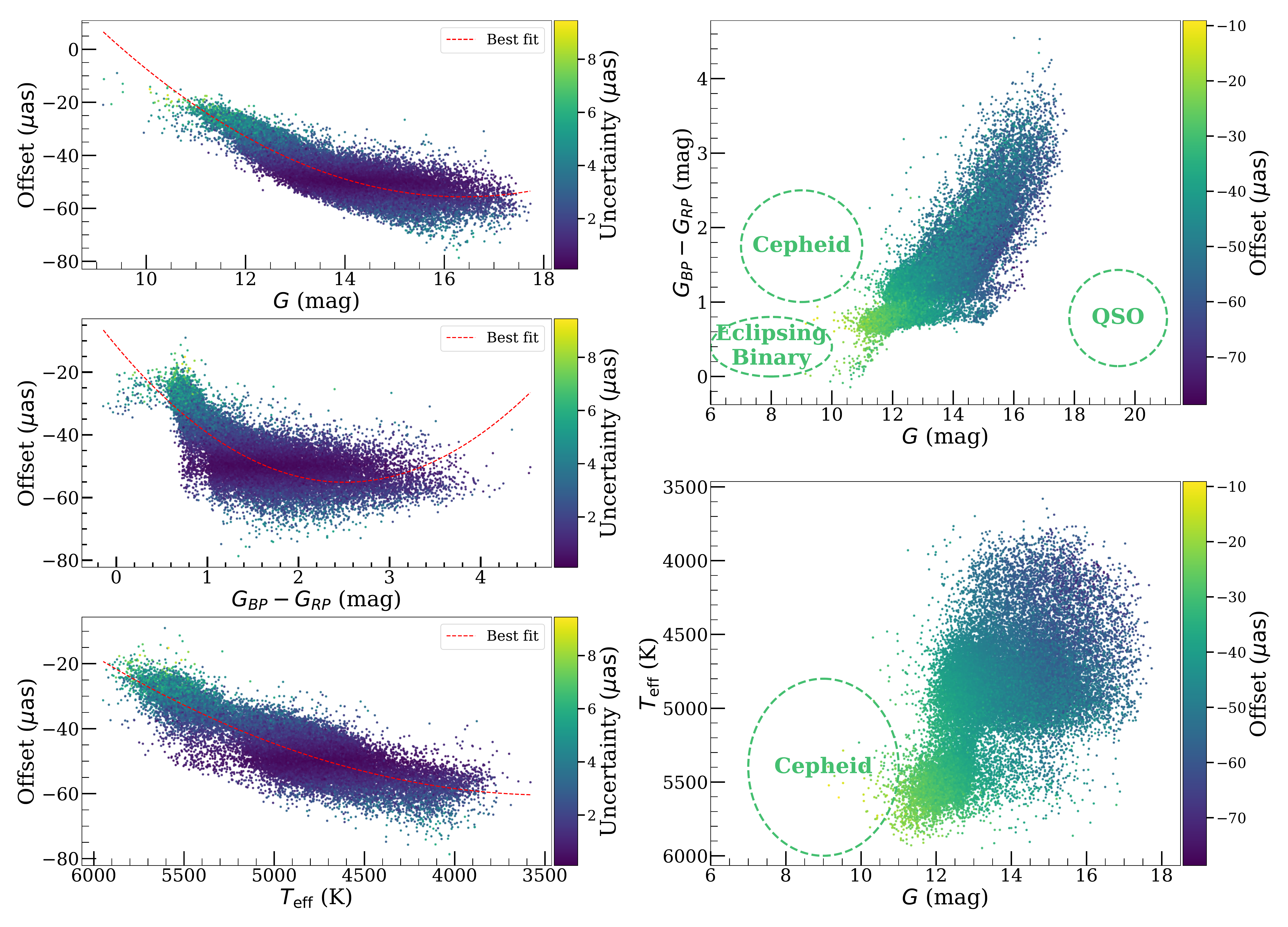}
\caption{The offset determined using the offset calibration model shown in \figurename~\ref{figure:nn_flow}, in which the zero-point offset depends on $G$, $G_{BP} - G_{RP}$, and \teff. The left panels show the multivariate offset being projected as a function of one of $G$, $G_{BP} - G_{RP}$, and \teff; the points are color-coded by the zero-point uncertainty which is estimated by dropout variational inference. There is a clear horizontal band of low zero-point uncertainty located at $\approx 50\,\uas$ which is similar to the zero-point offset that we determine with the constant offset model. The right panel displays two-dimensional projections of the three-dimensional offset model. The rough location of quasars (QSO) used by \citet{2018A&A...616A...2L}, Cepheids used by \citet{2018ApJ...861..126R}, and eclipsing binaries used by \citet{2019arXiv190200589G} are represented by ellipses with sizes that correspond to the typical range of the data used in these works, which all find an offset of $\approx -30\,\uas$. The \gaia\ DR2 zero-point offset becomes bigger in magnitude for fainter, redder, and cooler stars.}
\label{figure:offset}
\centering
\end{figure*}

\begin{figure}
\centering
\includegraphics[width=0.5\textwidth]{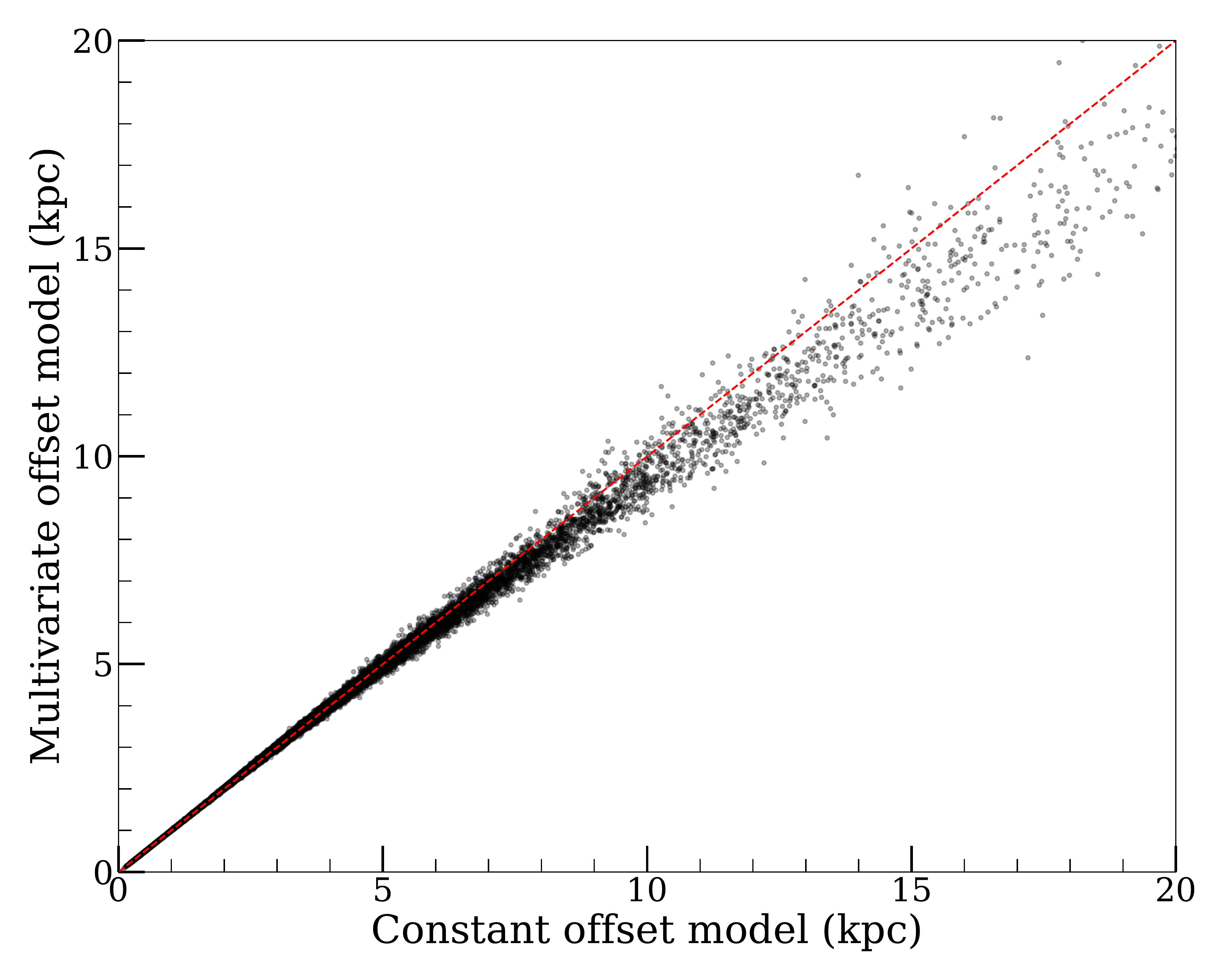}
\caption{Comparison between the NN distances for the APOGEE test set obtained with the constant parallax zero-point offset calibration model and those determined using the multivariate offset model. Both methods agree well within 10 kpc from the Sun. Beyond that, even a slight difference in offset leads to an overall offset and a greater dispersion. The multivariate offset distances are smaller at large distances, because the multivariate zero-point offset is such that \gaia\ underestimates the parallax by a greater amount for fainter stars.}
\label{figure:offset_comparison}
\centering
\end{figure}

We train models to calibrate the \gaia\ DR2 zero-point offset in 3 different variations while training the spectro-photometric NN to infer luminosity from spectra. We train a first model on unmodified \gaia\ DR2 parallaxes without any offset calibration model to investigate the spectro-photometric distances that we obtain without calibrating or correcting the zero-point offset. We train a second model on unmodified parallaxes with a constant zero-point offset calibration (see Section~\ref{subsec:models} for implementation details). The third model that we fit is trained  on unmodified parallax with a zero-point offset calibration model that depends on $G$, $G_{BP} - G_{RP}$, and \teff. 

Comparing the first model that does not calibrate or correct the zero-point offset with the constant-offset second model demonstrates clearly that there is indeed a zero-point offset in \gaia\ DR2. This is evident from \figurename~\ref{figure:offset}, where we compare the parallax obtained from the spectro-photometric NN with no zero-point offset to the parallax obtained when fitting a constant zero-point offset. Because there is a zero-point offset, the spectro-photometric NN parallax trained without accounting for the offset is unable to match the \gaia\ parallax over the entire range of parallaxes. Because the training sample is dominated by distant giants, the NN optimizer matches the NN parallaxes to the \gaia\ parallaxes at small parallax, but fails to do so for similar stars at larger parallaxes, leading to an increasing offset between NN and \gaia\ parallaxes at larger parallaxes. The model that includes a fitted constant zero-point offset is much better at matching the \gaia\ parallax over a wide range of parallaxes; the fact that it does not do so perfectly is because the zero-point offset is not constant, as we discuss further below.

When fitting for a constant zero-point offset, we get a zero-point offset of $-52.3\pm 2.0 \,\uas$; these values are obtained by inputting featureless vectors of ones in the offset calibration model and sampling the posterior of the offset calibration model by dropout to get the result and uncertainty (see Section~\ref{subsec:models}). This value is similar to that found by \citet{2018arXiv180502650Z} also using red giant stars. Our result deviates from the $\approx -30 \,\uas$ determined with quasars \citep{2018A&A...616A...2L}; this indicates that the zero-point may be different for different types of stars and for objects with different spectral energy distributions such as quasars (as indicated by their color or effective temperature).

To investigate whether the zero-point offset depends on other properties, we fit the multivariate offset model that depends on $G$, $G_{BP} - G_{RP}$, and \teff. This model therefore determines an individual zero-point offset for each star and we determine the zero-point and its uncertainty for each star by sampling the posterior of the offset calibration model using dropout variational inference. Taking the mean of the zero-point offsets in our testing sample, we obtain $\approx-50\,\uas$ in good agreement with our fit of a constant zero-point offset above. The full dependence on different properties in the testing sample of the zero-point offset is displayed in \figurename~\ref{figure:offset}. The left column shows the zero-point as a function of one of the properties ($G$, $G_{BP} - G_{RP}$, and \teff) at a time and the right column displays the zero-point offset as a function of $G$ and $G_{BP} - G_{RP}$ and of $G$ and \teff. The points in the left column are color-coded by the uncertainty in the zero-point offset and it is interesting to notice that a horizontal band of low uncertainty is located at $\approx 50\,\uas$, which is similar to the result we have estimated with the constant offset model. The majority of the stars in the APOGEE red-giant sample have magnitudes, colors, and temperatures such that they fall in this $\approx 50\,\uas$ low-uncertainty regime.

From the trends in the left column of \figurename~\ref{figure:offset}, the offset seems to be increasing in magnitude with $G_{BP} - G_{RP}$ and $G$, i.e., apparently redder and fainter stars seem to have more negative offset. The offset is also smaller with increasing surface temperature, consistent with the color trend. This behavior might explain why \citet{2018arXiv180502650Z} reports that the offset seems to have a dependence on parallax that is such that the zero-point offset is larger when the parallax is smaller: in a sample of giants like the APOKASC sample used by \citet{2018arXiv180502650Z}, stars farther away are generally fainter, and thus have a larger offset due to the strong dependency of the zero-point offset on $G$ that we find. For RC stars, \citet{2018arXiv180502650Z} reports $-50.2 \pm 2.5\text{(stat.)} \pm 1\text{(syst.)}\,\uas$ while our inverse-variance weighted mean of offset for the same RC stars is $\approx -48\,\uas$ with negligible uncertainty. \citet{2018arXiv180502650Z} attribute the different offset they find for RGB and RC stars to a systematic in the asteroseismic radius scale, but given that we confirm this difference without relying on the radius scale, it appears that this is instead a true difference in parallax zero-point offset for RC and RGB stars.

Similarly, the trends that we find might also explain the much smaller offset found by \citet{2018A&A...616A...2L} using quasars, because quasars are much bluer and effectively hotter and the color and \teff\ trends that we find are such that these properties should lead to a smaller offset. However, quasars are also generally fainter than the stars in our sample and the $G$ trend that we find would point towards a \emph{larger} offset. The color-magnitude-temperature range of quasars is far from the range covered in this work with giants and so a direct comparison between our zero-point offsets and those found using quasars is difficult.

{\allowdisplaybreaks
We summarize the results from \figurename~\ref{figure:offset} by providing simply polynomial models of the univariate trends in the left column panels. These are obtained using quadratic regression and we find (we do not provide the statistical uncertainty in these fits as it is negligible):

\begin{align}
\plx_\mathrm{offset} /\uas&=1.18\ (G-16)^2 -92\ (G-16) - 55.49\\
\phantom{\plx_\mathrm{offset} /\uas}& \quad (11 \lesssim G \lesssim 17)\,,\nonumber\\
\plx_\mathrm{offset} /\uas&=6.90\ (G_{BP} - G_{RP})^2 - 34.63\ (G_{BP} - G_{RP})\\
& \ \  - 11.67 \quad (0.8 \lesssim G_{BP} - G_{RP} \lesssim 3)\,,\nonumber\\
\plx_\mathrm{offset} /\uas&=6\times10^{-6}\ (\teff/\mathrm{K} - 4500)^2 + 0.0348\ (\teff/\mathrm{K} - 4500)\nonumber\\
& \ \  - 53.17 \quad (4000\,\mathrm{K} \lesssim \teff \lesssim 5750\,\mathrm{K})\,.
\label{eq:offset3}
\end{align}
}

Our zero-point estimates can be applied to \emph{all} 139,847,389 stars in \gaia\ DR2 within the color--magnitude range covered by APOGEE.

\section{Results: Precision and Accuracy of spectro-photometric distances}\label{sec:precision}

\begin{figure}
\centering
\includegraphics[width=0.5\textwidth]{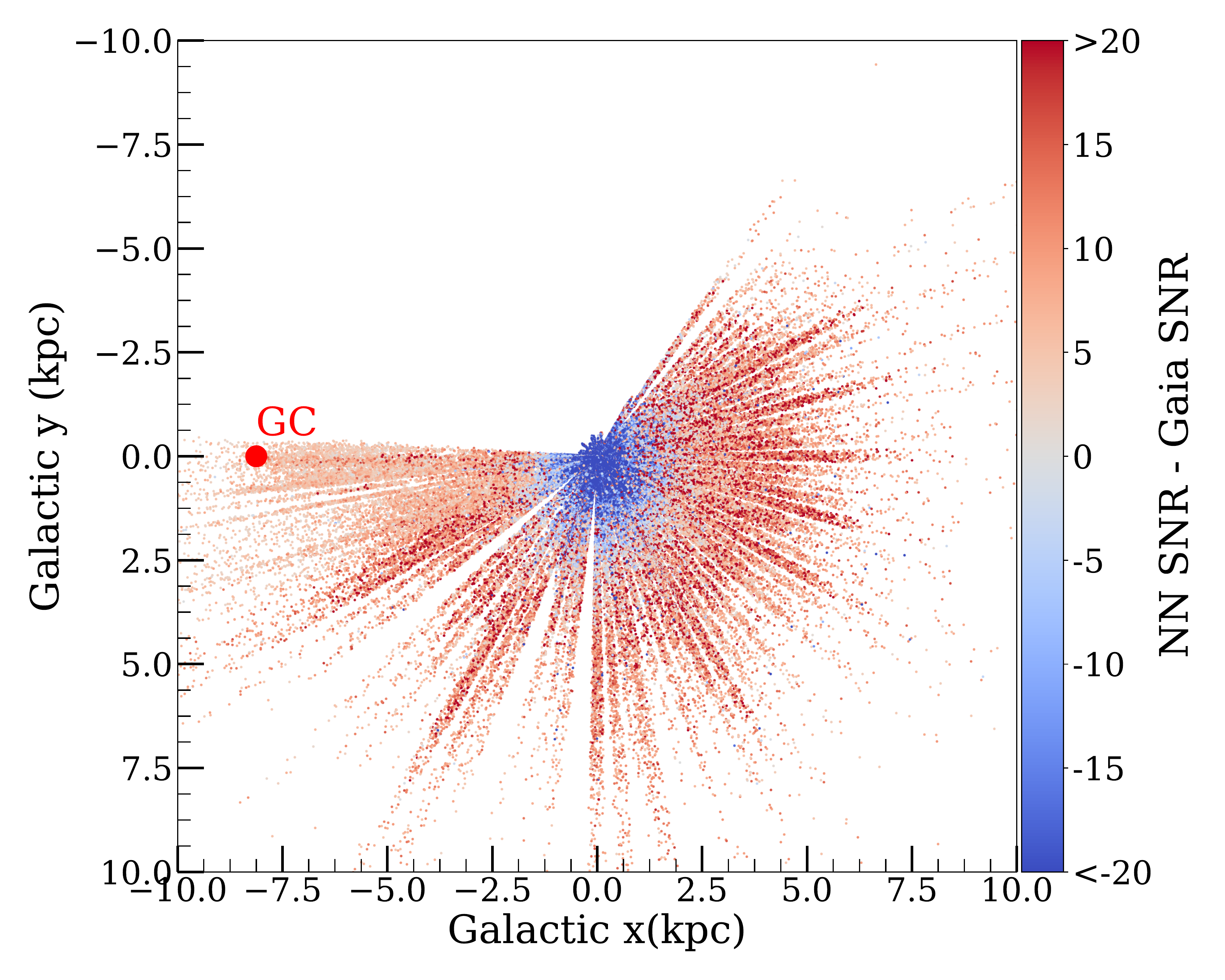}
\caption{Signal-to-Noise Ratio (SNR) difference between NN and \gaia\ distances displayed in Galactic x and y coordinates for stars within 500\,pc from the Galactic mid-plane. Blue colors indicates that \gaia\ has a higher SNR distance than the NN while red colors indicate that the NN distance is better. The transition from blue to red happens $\approx 2\,\mathrm{kpc}$ from the Sun, meaning that adopting the NN distance at $\gtrsim 2\,\mathrm{kpc}$ gives a better distance.}
\label{figure:snr_xymap}
\centering
\end{figure}

\begin{figure*}
\centering
\includegraphics[width=\textwidth]{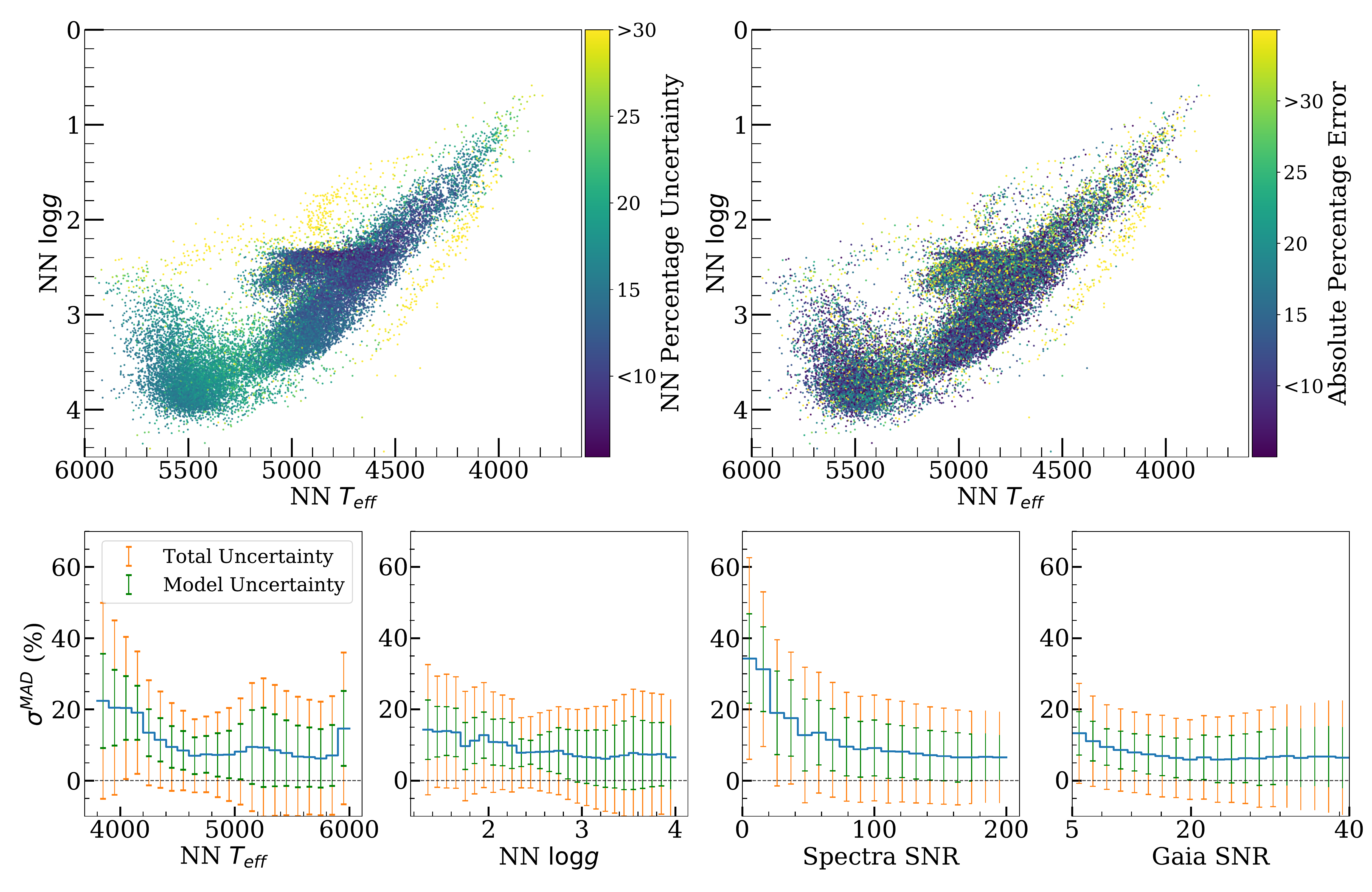}
\caption{Precision and accuracy of the spectro-photometric NN parallaxes in the test set. The top two panels show the percentage uncertainty of the NN parallaxes $\plx_{\mathrm{NN}}$ (top left) and the absolute percentage error between $\plx_{\mathrm{NN}}$ and $\plx_{\mathrm{Gaia}}+52\,\uas$ (top right) as a function of temperature and surface gravity determined by the NN of \citet{2019MNRAS.483.3255L}. The percentage uncertainty returned by the NN model is correlated with the actual error, i.e., the absolute percentage error is generally larger for uncertain predictions. The four bottom panels display the median absolute percentage deviation $\sigma^{\mathrm{MAD}}$ (see \eqnname~\eqref{eq:scattering}) as a function of \teff, \logg, spectral SNR, and SNR of the \gaia\ parallax measurement; the error bars are the median uncertainty for the NN parallax predictions in each bin (the green error bar is the model uncertainty component, while the orange error bar is the total uncertainty). The prediction is generally accurate to $<8\%$, except for the region where $\teff<4250\,\mathrm{K}$ or $\logg<2$, because training data is sparse in this region of luminous giants. The NN parallax precision starts to significantly degrade for spectral SNR $< 50$. The NN parallax precision is constant with \gaia\ parallax SNR and, at spectral SNR $> 100$, with  spectral SNR, demonstrating that the NN parallax uncertainty is due to scatter in the relation between spectra and luminosity, which fundamentally limits the distance precision that can be obtained from APOGEE spectra.}
\label{figure:precision}
\centering
\end{figure*}

\subsection{Comparison to \emph{Gaia}}

After we train the spectro-photometric NN on high SNR APOGEE spectra, we test the model on the low SNR spectra test set. To make our results easier to compare to other approaches, we use the constant zero-point offset model for this comparison---the discussion in Sec.~\ref{sec:offset_result} demonstrates that for the bulk of the APOGEE stars the constant zero-point offset model works about as well as the more complicated ($G$, $G_{BP} - G_{RP}$,  \teff)-dependent offset model. \figurename~\ref{figure:offset_comparison} shows a comparison between the distances obtained using the constant and the multivariate offset models for stars  in the test set. Both models agree well within 10 kpc from the Sun, but beyond that even a slight difference in the zero-point of a few \uas\ in parallax results in a noticable offset and a greater dispersion. Our multivariate zero-point model suggests that fainter stars have bigger offsets (in absolute value) and, therefore, stars beyond 10 kpc are generally closer when using the multivariate offset model than when using the constant zero-point model. The bulk of the APOGEE stars are within 10 kpc and we therefore use the constant zero-point model when evaluating how well our NN distances perform.

We test the parallaxes returned by the spectro-photometric NN in combination with the extinction-corrected $K_s$ magnitudes by comparing them to the \gaia\ parallaxes, correcting the latter as $\plx_\mathrm{Gaia}+52\,\uas$ to account for the zero-point offset. For APOGEE DR14 as a whole, \figurename~\ref{figure:snr_xymap} compares the SNR of the distance determined using the NN to that determined by \gaia\ as a function of Galactic x and y coordinates. The blue scatter points, where \gaia\ performs better than the NN, are concentrated at $\lesssim 2\,\mathrm{kpc}$ from the Sun. Beyond 2 kpc, the NN distances are better than the \gaia\ distances. The bulk of the APOGEE giants are at distances greater than 2 kpc, so the NN distances outperform \gaia\ for most APOGEE giants.

A summary of the comparison to \gaia\ parallaxes with uncertainty $<20\%$ in the test set is displayed in \figurename~\ref{figure:precision}. The top panels show the percentage uncertainty returned by the NN---this is the combination of the model and predictive uncertainty --- on the left and the absolute percentage error between $\plx_{\mathrm{NN}}$ and $\plx_{\mathrm{Gaia}}+52\,\uas$ on the right. The absolute percentage error is generally larger for predictions with a more uncertain prediction, showing that the uncertainty in the NN parallax determined by the model is reasonable. The bottom panels show the median of the  absolute percentage deviation (MAD \%) as a function of \teff, \logg, spectral SNR, and $\plx_{\mathrm{Gaia}}$ SNR. The prediction is generally accurate to $<8\%$ all the way from dwarfs with solar luminosity to giants brighter than the red clump. For the most luminous giants with $\teff<4250\,\mathrm{K}$ or $\logg<1.6$, the parallax precision is worse at around 20\% or higher. The error bars indicate the typical uncertainty in the NN parallax prediction and the fact that the error bars typically reach zero and not much further demonstrates that the uncertainty estimates returned by the NN are a reasonable description of the precision in the NN parallax.

The two remaining lower panels of \figurename~\ref{figure:precision} show the $\sigma^{\mathrm{MAD}}$ \% absolute deviation as a function of the SNR in the APOGEE spectra or in the \gaia\ parallaxes used in the comparison. The NN parallax precision is constant for all but the lowest \gaia\ parallax SNR values, indicating that the magnitude of the $\plx_\mathrm{NN}-\plx_\mathrm{Gaia}$ deviation is driven by noise in the NN parallax rather than in the \gaia\ comparison parallax for these high-SNR \gaia\ parallaxes. For spectra with spectral SNR $>100$ the NN parallax precision is roughly constant, which shows that the parallax precision is not limited by noise in the spectra at high spectral SNR, but rather by the scatter in the spectra--luminosity relation. Below spectral SNR of 100, the precision in the NN parallax degrades and especially so for spectral SNR $< 50$. Thus, if future high-resolution spectroscopic surveys in the $H$ band like SDSS-V \citep{2017arXiv171103234K} want high-precision spectro-photometric distances for luminous giants, these surveys should aim for SNR at least 50 and ideally 100. 

\begin{figure}
\centering
\includegraphics[width=0.5\textwidth]{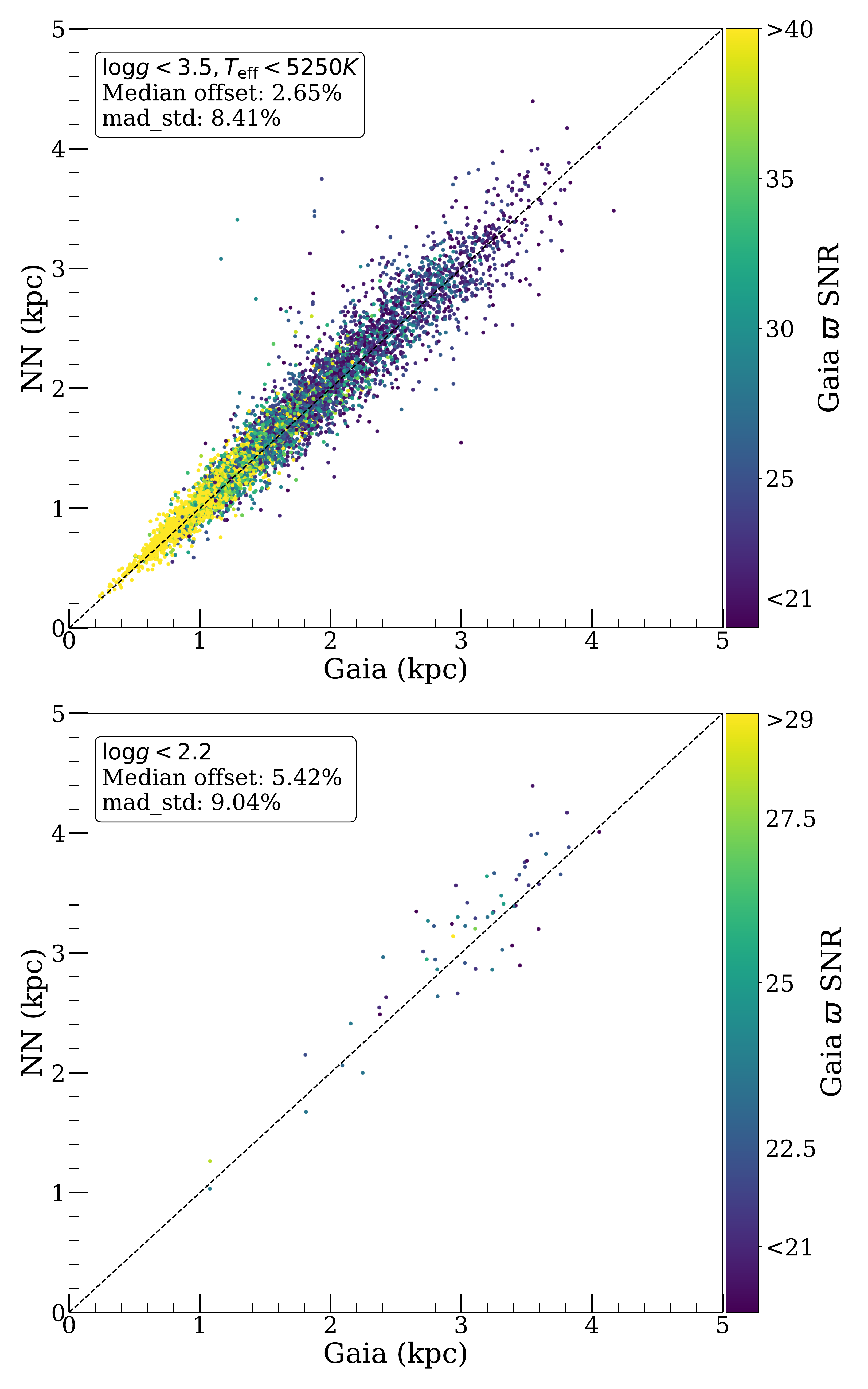}
\caption{Comparison between $1/\plx_{\mathrm{NN}}$ and $1/[\plx_{\mathrm{Gaia}}+52\,\uas]$ for stars with a relative uncertainty in $ \plx_{\mathrm{Gaia}}+52\uas$ less than 5\%. The top panel shows the comparison for the full APOGEE giant sample with such good \gaia\ parallaxes ($\logg<3.5$ and $\teff<5250\,\mathrm{K}$), while the bottom panel focuses on the most luminous giants by additionally requiring $\logg < 2.2$. The NN parallaxes are precise to about $8$ to $9\%$ and highly accurate, even at large distances.}
\label{figure:gaia_5}
\centering
\end{figure}

To further validate the accuracy of the NN distances, \figurename~\ref{figure:gaia_5} compares $\plx_{\mathrm{NN}}$ and $\plx_{\mathrm{Gaia}}+52\,\uas$. The \madstd, is used a robust measurement of the scatter which is based on the Median Absolute Deviation (MAD): $\madstd = 1.4826\,\mathrm{MAD}$, where the factor is such that for a Gaussian distribution $\madstd$ equals the Gaussian standard deviation. Thus, for a set of percentage residual $R\%:[R_1\%, R_2\%,...,R_n\%]$ which is 100\% times residual $R$ over ground truth, \madstd\% is

\begin{equation}  \label{eq:scattering}
\madstd\% = 1.4826\,\mathrm{median}\left(|R_i\% - \mathrm{median}(R\%)|\right)\,.
\end{equation} 

To focus the comparison on high-quality \gaia\ inverse parallaxes we only display stars with $\plx_{\mathrm{Gaia}}+52\,\uas$ uncertainty that is lower than 5\% in the test set, the \madstd is $\approx 8\%$ when considering all giants (top panel) and only slightly larger at $\approx 9\%$ for the most luminous giants (lower panel). The median difference between the NN and the \gaia\ distance is only $\approx2.5\%$ for all giants and about twice as large for the most luminous giants. Thus, our NN distances are highly accurate. Note that because essentially all giants have $\plx < 1\,\mas$ and the \gaia\ DR2 zero-point offset has magnitude and color-dependent (and perhaps spatial) trends at the level of tens of $\uas$ (see Sec.~\ref{sec:offset_result}), the precise accuracy is difficult to determine because we cannot trust the \gaia\ parallaxes at the percent-level for these distances.

\subsection{Comparison to other spectro-photometric distances}\label{subsec:comparison}

\begin{figure*}
\centering
\includegraphics[width=\textwidth]{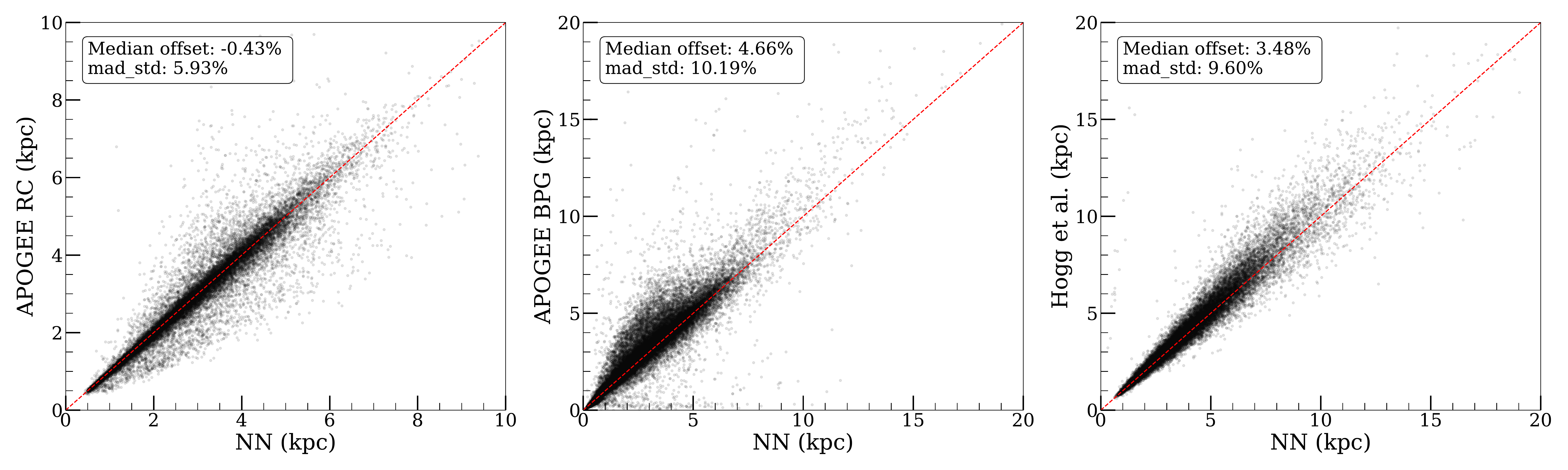}
\caption{Comparison of the NN distances from this work to three different spectro-photometric distances: APOGEE red-clump giants (\citealt{2014ApJ...790..127B}; left panel), APOGEE BPG distances (\citealt{2016A&A...585A..42S}; middle panel), and distances for luminous red giants from (\citealt{2018arXiv181009468H}; right panel). We cut the sample to those stars with  $<20\%$ uncertainty in both distances that are compared in each panel.}
\label{figure:comparison}
\centering
\end{figure*}

To further test the performance of the NN parallaxes, we compare our results to previous spectro-photometric distance estimates for subsets of the APOGEE data: those for RC stars from the APOGEE-RC catalog \citep{2014ApJ...790..127B}, distances from the APOGEE-DR14 Brazilian Participation Group (BPG) Distance Estimation Catalog \citep{2016A&A...585A..42S}, and the distances for luminous red giants from \citet{2018arXiv181009468H}.

The DR14 version of the APOGEE-RC catalog \citep{2014ApJ...790..127B} contains 29,502 stars with $\approx95\%$ purity and with distances precise to $5\%$ to $10\%$. RC stars are stars in the core Helium-burning stage in the stellar evolution of low-mass stars and because they have a narrow luminosity distribution they act close to a standard candle. The distances in the RC catalog are determined using predictions of their absolute magnitude from stellar evolution models. The overall distance scale in the RC catalog was calibrated against parallaxes in \emph{Hipparcos} and therefore does not have the same zero-point offset issue as the \gaia\ DR2 data. Thus, the RC catalog can provide precise and accurate distances to test both the NN precision as well as the zero-point offset correction. We remove 1,408 RC stars from consideration because they are likely contaminants: these stars have a difference between the \logg\ value determined by the APOGEE pipeline and the data-driven NN approach of \cite{2019MNRAS.483.3255L} that is larger than 0.2 dex. Because the APOGEE RC catalog selection is based on the pipeline value of \logg\ and we believe the NN \logg\ value to be more accurate, these 1,408 stars are likely mislabeled.

The catalog of APOGEE BPG distances that is included in APOGEE DR14 contains 211,243 stars. Distances to these stars are determined using a Bayesian method applied to spectroscopic parameters and photometry that makes use of stellar evolutionary models and a Galactic density prior based on mass, age, and metallicity.  Distances have uncertainties of  $\approx10\%$ for dwarf stars and $\approx20\%$ for giants.

The distances for luminous red giants from \citet{2018arXiv181009468H} are the only ones among the three comparison distances that are determined using purely data-driven techniques that are similar to those used in this work. \citet{2018arXiv181009468H} employ a linear (in exponential space) model for how parallax depends on the APOGEE spectrum and broad-band photometry. Their model has significantly fewer trainable parameters than the NN used in this work. \citet{2018arXiv181009468H} only provide distances for the APOGEE luminous red-giants with \logg < 2.2, $(J-K)<(0.4\text{mag})+0.45(G_{BP}-G_{RP})$ and ($(H-W_2)>(-0.05\text{mag})$ where $J,\ K,\ H,\ W_2$ are $2MASS$ and $WISE$ photometry. \citet{2018arXiv181009468H} report 10 percent distance estimates.

The distances determined by the NN from this work are compared to those for the same stars in the three comparison catalogs in \figurename~\ref{figure:comparison}. The median offset and the scatter in the offset are indicated in the legend of each panel. Overall, there is almost no bias between the NN distances and those in the APOGEE-RC catalog (median offset is $<1\%$); the scatter in the offset is $\approx6\%$, similar to the precision quoted for RC stars by \citet{2014ApJ...790..127B}. The typical NN uncertainty for RC stars is $\approx 10\%$. Thus, the NN distances for RC stars are very similar to those in the RC catalog, even though the distance estimates are obtained using very different methods.

The middle panel of \figurename~\ref{figure:comparison} compares our NN distances to those in the APOGEE BPG distance catalog. Overall, there is about a $5\%$ offset in the sense that BPG distances are larger than those returned by the NN. This trend is driven by stars located toward the center of the Galaxy and is likely caused by the Galactic-density prior used by the BPG method. This prior contains a massive bulge that has the effect of making it more likely that a star is deep within the bulge. For stars located in the direction of the outer Galaxy, the NN and BPG distances agree much better. Aside from this overall shift toward larger distances in the BPG catalog, the robust scatter between the NN and BPG distances is just over $10\%$, but note that there are many stars in the tails of the difference distribution.

The right panel of \figurename~\ref{figure:comparison} compares our results to the distances from \citet{2018arXiv181009468H}. Overall, there is only a small $\approx3\%$ offset between these two distance estimates with scatter $\approx10\%$ and few outliers. The good agreement between our and the  \citet{2018arXiv181009468H} distances holds over the entire $\approx 10\,\mathrm{kpc}$ distance range contained within the luminous red giant sample. The uncertainties in the distances reported by \citet{2018arXiv181009468H} are likewise similar to our NN model uncertainty. This is a remarkable agreement between these two different methods. 

\section{Abundances across the Milky Way}\label{sec:MW_chem}

To illustrate the power of the NN distances determined in this work, we make maps over a large fraction of the Milky Way disk of elemental abundance ratios for the elements measured by APOGEE. To do this, we combine the NN distances from this work with the NN-determined stellar parameters and abundances from \citet{2019MNRAS.483.3255L}. We select stars from APOGEE DR14 with NN distance uncertainty less than 20\%, \logg\ uncertainty less than $0.2\,\mathrm{dex}$ (as explained in \citealt{2019MNRAS.483.3255L}, this cut removes dwarfs for which the APOGEE and NN abundance measurements are unreliable), and \xh{Fe} uncertainty less than $0.05\,\mathrm{dex}$. We convert three-dimensional coordinates from the heliocentric to the Galactocentric coordinate frame by adopting the distance to the Galactic center of $8.125\,\mathrm{kpc}$ from \citet{2018A&A...615L..15G} and the Sun's height above the plane of $20.8\,\mathrm{pc}$ from \citet{2019MNRAS.482.1417B}. To focus on trends in the abundance ratios within the disk, we further require that stars be within $300\,\mathrm{pc}$ from the Galactic mid-plane; 52,476 stars pass all of these cuts. When we consider abundance ratios relative to iron, we additionally only use stars with \xh{X} uncertainty less than $0.05\,\mathrm{dex}$ for the considered element; this creates samples ranging in size from 15,833 (for Na) to 52,476 (for Si and Ni).

\begin{figure}
\centering
\includegraphics[width=0.5\textwidth]{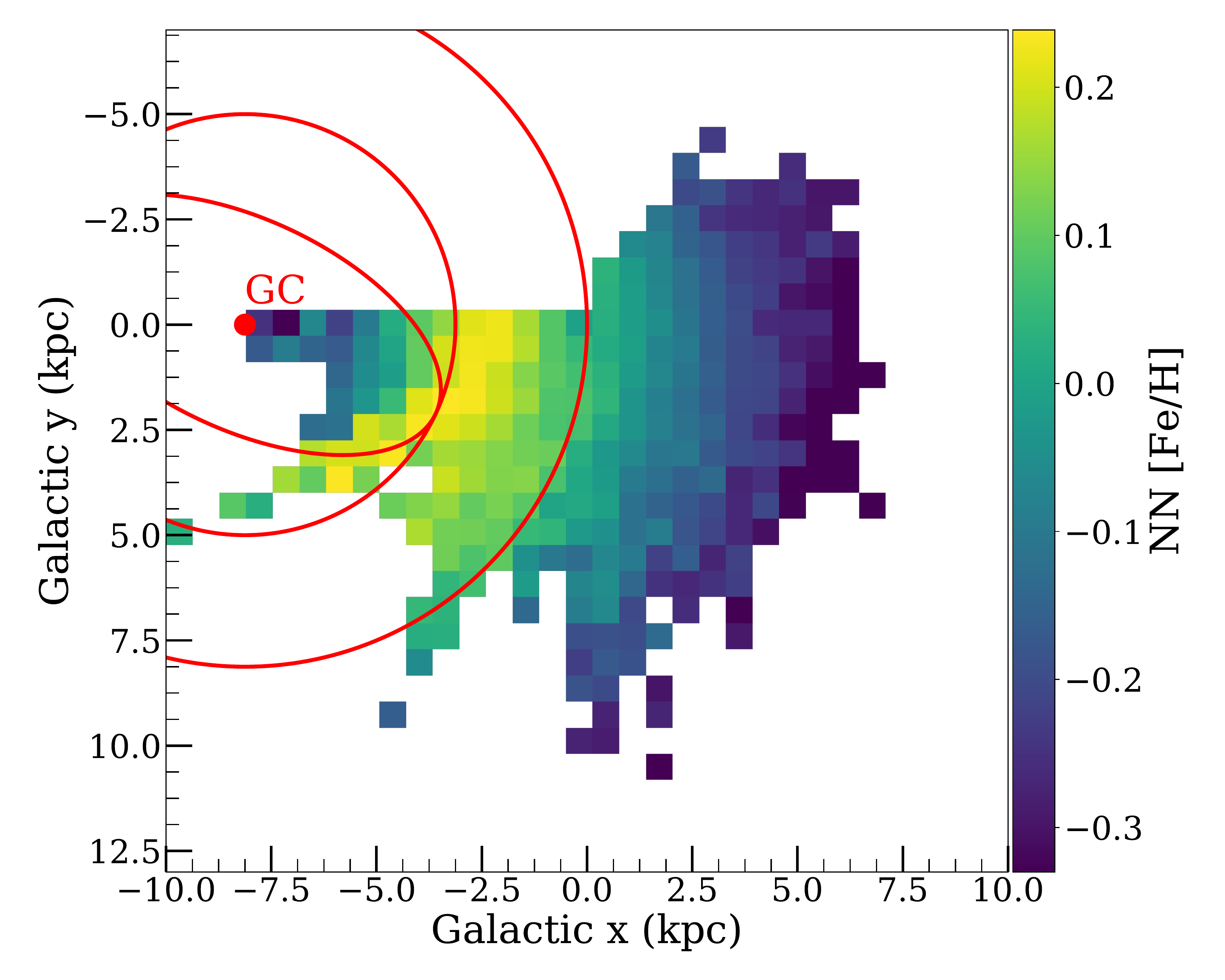}
\caption{\xh{Fe} as a function of position in the Galactic disk for stars within 300 pc from the disk's mid-plane. The map displays the median \xh{Fe} in bins $\approx (500\,\mathrm{pc})^2$ in size across the Milky Way disk using NN abundances and distances. Each bin contains at least 10 stars. GC denotes the Galactic Center. The two circles centered on GC represent $5\,\mathrm{kpc}$ and $8.125\,\mathrm{kpc}$ (the adopted Solar circle). The ellipse centered at GC illustrates  a $5\,\mathrm{kpc}$ bar with a 2:1 axis ratio oriented at 25 degree. \xh{Fe} in the Milky Way peak at $R\approx5\,\mathrm{kpc}$, likely due to the transition to the barred region, and decline strongly from this peak toward the Galactic center and towards the outer disk.}
\label{figure:mw_chem}
\centering
\end{figure}

\begin{figure*}
\centering
\includegraphics[width=\textwidth]{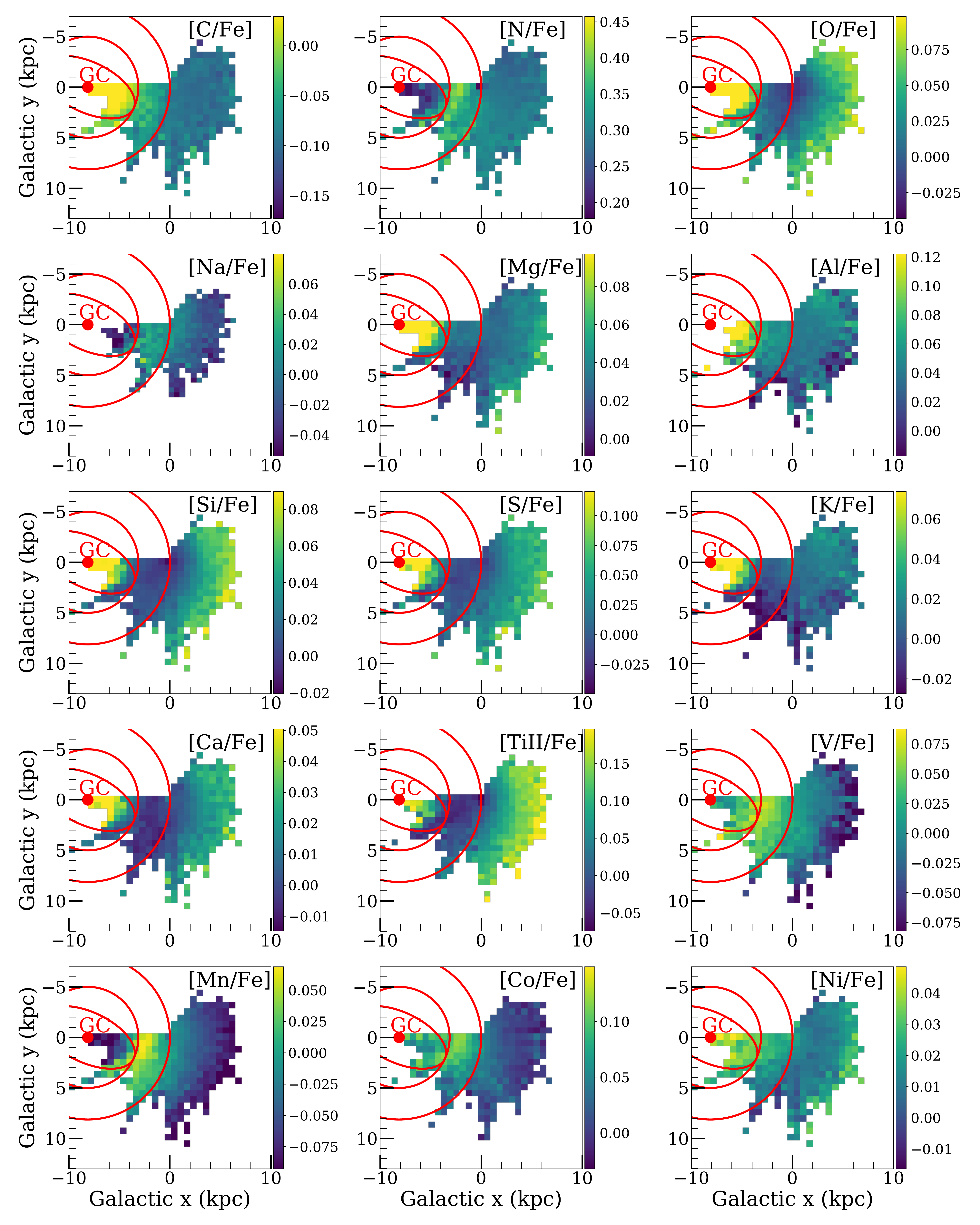}
\caption{Maps of \xfe{X} abundance ratios for 15 different elements measured from APOGEE spectra across the Galactic disk. As in \figurename~\ref{figure:mw_chem}, the map displays the median \xfe{X} in bins $\approx (500\,\mathrm{pc})^2$ in size for bins with more than 10 stars. The circles, ellipse, and other markings are as in \figurename~\ref{figure:mw_chem}.}
\label{figure:all_elem}
\centering
\end{figure*}

\figurename~\ref{figure:mw_chem} and \figurename~\ref{figure:all_elem} show the median elemental abundance ratios in bins of $\approx (500\,\mathrm{pc})^2$ size across the Milky Way disk. It is clear that \xh{Fe} peaks at a Galactocentric radius of $\approx 5\,\mathrm{kpc}$ and that \xh{Fe} decreases monotonically inwards and outwards from this peak. This is markedly different from the behavior found in the inner Galaxy by \citet{Hayden14a} using APOGEE DR10 data, which showed a flattening of the metallicity gradient within $\approx 6\,\mathrm{kpc}$, but with a slight increase of the metallicity towards the center all the way to $R \approx 2\,\mathrm{kpc}$. This difference is most likely due to the adopted distances. \citet{Hayden14a} determined distances using a Bayesian methodology using a Galactic density prior similar to that of the BPG distances that we compared to in Section~\ref{subsec:comparison} and the effect of this prior is to place stars towards the bulge at larger distances. Because of the good agreement between our distances and the \gaia\ parallaxes for bright, luminous giants on the one hand and the alternative distance determination from \citet{2018arXiv181009468H} on the other hand, we believe that our distances for stars towards the bulge are more reliable. Thus, it appears that \xh{Fe} in the Milky Way peaks at $\approx 5\,\mathrm{kpc}$ from the center and declines to $\approx -0.3\,\mathrm{dex}$ in the center. That the bulge/bar region is more metal-poor than the innermost reaches of the disk is in agreement with other analyses of ARGOS and APOGEE data \citep{2013MNRAS.430..836N,2016ApJ...819....2N}, but our larger sample and precise distances allow the spatial metallicity trend in the inner Milky Way to be mapped in much greater detail.

\figurename~\ref{figure:all_elem} displays the spatial trend in elemental abundance ratios with respect to iron over a wide area of the Galactic disk. Abundance ratios of alpha elements (\xfe{O}, \xfe{Mg}, \xfe{Si}, \xfe{S}, \xfe{Ca}, and \xfe{TiII}\footnote{We use the Ti abundance determined only from the TiII line in the APOGEE spectral region, due to issues with the analysis of neutral Ti lines in the APOGEE analysis \citep[e.g.,][]{2016A&A...594A..43H}}) largely trace each other in all parts of the Galactic disk, although notably \xfe{Mg} stays relatively constant towards larger Galactocentric radii, while other alpha elements like \xfe{O}, \xfe{Si}, and \xfe{Ti} display a clear upturn towards the outer disk. Odd-Z elements \xfe{Al} and \xfe{K} trace the alpha element \xfe{Mg} everywhere. Among iron-peak elements, \xfe{Ni} is almost constant everywhere, as expected, while \xfe{V}, \xfe{Mn}, and \xfe{Co} show a peak at $\approx 5\,\mathrm{kpc}$ similar to that in the \xh{Fe} map. \xfe{C} and \xfe{N} are approximately constant outside the solar circle, but inside the solar circle and especially inside the bulge/bar, \xfe{C} strongly increases and is anti-correlated with \xfe{N}, which decreases. The spatial abundance ratio trends place strong constraints on nucleosynthetic yields from different processes and on the history of chemical enrichment in different parts of the disk that we will pursue in future work.

Finally, it is interesting to note that the shape of constant abundance-ratio contours in the inner Galaxy does not appear to be Galactocentric circles, but that they appear to be elliptical and aligned with the bar in the inner Milky Way (illustrated with the red ellipse in \figurename s~\ref{figure:mw_chem} and \ref{figure:all_elem}). Thus, we tentatively detect the bar in the spatial behavior of elemental abundance ratios. Previous determinations of the bar's shape are based on star counts only \citep[e.g.,][]{1991ApJ...379..631B,2013MNRAS.435.1874W,2015MNRAS.450.4050W} and the bar is also detected clearly in the kinematics of stars and gas in the inner Milky Way \citep[e.g.,][]{1991MNRAS.252..210B,2017MNRAS.465.1621P}. But our chemical Milky Way maps provide the first direct evidence from abundance ratios that the bar region extends out to 5 kpc, consistent with a fast-rotating, long bar as found by \citet{2017MNRAS.465.1621P}. Future data releases from APOGEE will include data from its southern extension, APOGEE-South, which will test this picture further by filling in the region at negative $y$ ($210^\circ \lesssim l \lesssim 360^\circ$) in the inner Milky Way.

\section{Discussion}\label{sec:discuess}

\subsection{Comparison to other \gaia\ DR2 zero-point offset determinations}

With the release of the \gaia\ DR2 data, \citet{2018A&A...616A...2L} provided a determination of the parallax zero-point offset of $-29\,\uas$ using quasars. Because the zero-point offset likely depends on magnitude, color, and sky location, subsequently a number of works attempted to estimate the zero-point offset.  \citet{2018arXiv180502650Z} used stars in the APOKASC sample of stars with asteroseismic and spectroscopic observations to estimate the offset as $-52.8 \pm 2.4\text{(stat.)} \pm 1\text{(syst.)\,\uas}$ for red-giant branch stars and $-50.2 \pm 2.5\text{(stat.)} \pm 1\text{(syst.)\,\uas}$ for RC stars, which as we discussed above is in good agreement with our determination of the zero-point offset. \citet{2018arXiv181009468H} similarly used luminous red giants in APOGEE to estimate the offset as $-48\,\uas$. \citet{2018ApJ...861..126R} used Milky-Way Cepheid variable stars using a previous joint determination of the period-luminosity relation and the distance ladder in the local Universe and estimated the offset to be $-46\pm13\,\uas$. Recently,  \citet{2019arXiv190200589G} employed Milky-Way detached eclipsing binary stars to obtain $-31\pm11\,\uas$.

The discrepancies between these different determinations of the zero-point offset are likely caused by the multivariate nature of the zero-point offset and we have determined the multivariate zero-point dependence using a flexible model (see \figurename~\ref{figure:offset}). \citet{2018arXiv180502650Z} and \citet{2018arXiv181009468H} both estimate similar offsets as we have found for the bulk of the red-giants stars used in this work, because these works use similar populations of red-giant stars in APOGEE as the majority of our sample. Both \citet{2018A&A...616A...2L} and \citet{2019arXiv190200589G} favor a smaller offset compared to our work, because (a) the quasars in \citet{2018A&A...616A...2L} are generally much bluer and dimmer compared to the giants in this work and (b) the eclipsing binaries used in \citet{2019arXiv190200589G} are bluer and brighter than the stars we consider. The quasars are so far outside of the magnitude and color range of stars in our sample and have such different spectral energy distributions that it is difficult to extrapolate our trends in \figurename~\ref{figure:offset} to quasars. The $(G,G_{BP}-G_{RP})$ panel in \figurename~\ref{figure:offset} demonstrates that we find that stars that are at the blue, bright end of our sample have offsets of $\gtrsim-30\,\uas$, similar to \citet{2019arXiv190200589G}. The uncertainty in the zero-point offset determined using Cepheids from \citet{2018ApJ...861..126R} is large enough to be consistent with all other determinations of the zero-point, including ours.

Compared to the methods employed by these other works, our technique does not rely on any physical or stellar model, but is instead entirely data-driven and our technique has far more freedom than the other methods. The determination of the parallax zero-point offset  is crucial to almost all applications of the \gaia\ DR2 data and it has significant implications for such questions as what the Hubble constant is.  The $(G,G_{BP}-G_{RP},\teff)$ dependence that we find in \figurename~\ref{figure:offset} points towards a zero-point offset for the bright, long-period Cepheids used in the determination of the Hubble constant that is smaller in absolute value and close to the value derived from quasars $-30\,\uas$. This would place Cepheids at somewhat larger distances than the distances used by \citet{2018ApJ...861..126R} and would therefore lead to a slight reduction in the inferred Hubble constant, reducing the tension with the value determined using cosmology and the inverse distance ladder \citep{2015PhRvD..92l3516A,2018arXiv180706209P}.

\subsection{Comparison to other spectro-photometric distances for red giants}

We compared the NN distances that we determined to three other methods for determining spectro-photometric distances in Sec.~\ref{subsec:comparison}. Here we further discuss the similarities and differences between our method and these other techniques:

\begin{itemize}
    \item Both the NN used in this work and the \citet{2018arXiv181009468H} technique map (APOGEE) spectra to luminosity in purely data-driven way, by training on the overlap between APOGEE and \gaia\ data. The BPG distance determinations rely on stellar models and a Galaxy density prior during inference. That we do not use a density prior or stellar models is both an advantage and a disadvantage. Our and the \citet{2018arXiv181009468H} distances are therefore solely informed by the data and they do not depend on imperfect stellar models and imperfect knowledge of the Galaxy's density distribution and how the APOGEE/\gaia\ selection function affects the observed number counts. On the other hand, we cannot easily extrapolate our results to stellar types outside of those included in our training set. But because we train directly on \gaia\ parallaxes (and simultaneously infer the \gaia\ zero-point offset), there is no doubt that our NN distances have higher precision and accuracy than those determined using stellar models and density priors, such as the BPG distances, \emph{for stars within the bounds of the training sample}.
    \item The \citet{2018arXiv181009468H} distances are determined using a much simpler model than the NN used in this work. They use a linear model for the logarithm of the parallax, whereas we use a multi-layer artificial NN as a universal function approximation. Because their model is so simple, \citet{2018arXiv181009468H} only train on bright red giants. The model in this work is complex and therefore it can infer luminosity for a wider range of stars. Our model returns precise distances, even at the edges of the training sample, and it also returns realistic uncertainties that can be used to identify stars for which the NN distance is unreliable (e.g., stars outside of the bounds of the training set). Fears that the complexity allowed by the NN leads to bad extrapolation are therefore unfounded for our model. We are therefore able to provide a single distance-estimation technique that can be applied from solar-luminosity dwarfs to the most luminous red giants
    \item The \citet{2018arXiv181009468H} distance methods contains a data-driven extinction model that makes use of additional broad-band photometry, while in our work we rely on external extinction data. This is not a great limitation for the application to APOGEE data, because extinctions determined using the RJCE method are highly reliable. But when applying our type of modeling to other data sets being able to include a data-driven extinction estimation would be highly desirable. It would be possible to construct a simpler NN to determine the extinction to each star that is similar to our existing offset calibration model, where in the extinction case the inputs would be multi-band broad-band photometric data similar to \citet{2018arXiv181009468H}.
    \item Different from \citet{2018arXiv181009468H} and essentially all other spectro-photometric distance methods, we calibrate the \gaia\ zero-point offset simultaneously with the training of the spectro-photometric model. We are even able to infer how the zero-point offset depends on stellar properties $G$, $G_{BP} - G_{RP}$, and \teff, with a flexible NN model for the zero-point offset. Thus, we have a consistent model for the \gaia\ DR2 zero-point offset and spectro-photometric distances.
\end{itemize}

\subsection{Future applications and challenges}

The method described in this work can be easily applied to future \gaia\ data release with the high level API of the \texttt{astroNN} package. The amount of data to train the algorithm will increase in future APOGEE and \gaia\ data releases. Other spectroscopic surveys such as GALAH \citep{2015MNRAS.449.2604D} and SDSS-V \citep{2017arXiv171103234K} will provide even larger training sets that also contain a wider range of stellar types. The flexibility of the NN technique will allow these data sets to be incorporated into a single method. Furthermore, future \gaia\ data releases will include low-resolution optical spectra for all \gaia\ sources. Training on these data will provide a way to improve the distance estimates of \emph{all} \gaia\ stars, although it remains to be determined by how much.

While the \gaia\ processing of the satellite measurements to determine parallaxes will improve and likely lead to smaller overall parallax zero-point offsets, it is likely that future data releases will still suffer from unknown zero-point issues. We have shown that the zero-point offset in \gaia\ DR2 is a function of $G$, $G_{BP} - G_{RP}$, and \teff\ and the zero-point may further depend on sky location (e.g., \citealt{2018A&A...616A...2L}) and other properties. We have shown that with the $\approx35,000$ member APOGEE training set, we can determine the zero-point offset to $\approx 2\,\uas$ precision, even if we allow the zero-point to depend on $G$, $G_{BP} - G_{RP}$, and \teff. Thus, simultaneously calibrating spectro-photometric distances and the multi-variate nature of the parallax zero-point is a way to obtain a high-precision zero-point calibration. However, determining the zero-point offset's multivariate dependencies is challenging, because the NN is complex. If we give the NN more degrees of freedom, the NN will may start deviating from physically plausible trends, as the NN only minimizes the objective function. For example, the zero-point offset is known to have spatial covariance and a quasi-regular triangular pattern with a period of about one degree (see the LMC test in \citealt{2018A&A...616A...2L}). We believe that it will be difficult for a NN, especially the Bayesian NN we employ here which is highly regularized, to capture such high-frequency periodic variations in equatorial coordinates, although it may be possible to recast the spatial dependence of the zero-point into a more well-behaved set of properties. 

All data-driven methods for determining the zero-point essentially suffer from the issue that the offset calibration might not be physical. Using standard(izable)-candle stars such as red clump stars or variable stars as well as using extra-galactic sources like quasars can only provide the offset calibration in the color/position/magnitude space covered by each sample. It is difficult to use an ensemble of them to average out the bias or inaccuracy introduced by each one. Because our method is flexible enough to allow for a wide range of stellar types to be incorporated simultaneously, it can determine the zero-point's dependence on color/position/magnitude consistently over a wide range in these properties.

\section{Conclusions}\label{sec:conclusion}

Milky Way stars are being surveyed astromatically, photometrically, and spectroscopically in an unprecedented manner thanks to survey like \gaia\ and APOGEE. The distance to each star is one of its most crucial properties and while \gaia\ is revolutionizing the measurement of stellar distances using its parallax measurements, \gaia\ can only determine precise distances within a small volume compared to the Galaxy as a whole. Existing survey like APOGEE---which will soon also release data from its southern extension, APOGEE-S---cover a wide volume of the Milky Way. Machine learning provides an opportunity to use the precise \gaia\ distances for nearby APOGEE stars to train a spectro-luminosity model and to use this model to provide precise spectro-photometric distances for a sample that covers a large part of the Milky Way disk.

In this paper, we have simultaneously calibrated spectro-photometric distances and the \gaia\ DR2 parallax zero-point offset with deep learning using neural networks. We have determined  spectro-photometric distances for the entire APOGEE DR14 sample of 277,371 stellar spectra which cover  Galactocentric radii $2\,\mathrm{kpc} \lesssim R \lesssim 19\,\mathrm{kpc}$. About 150,000 of these has $<10\%$ distance uncertainties. Our spectro-photometric distances have higher SNR than \gaia\ beyond $\approx2\,\mathrm{kpc}$. We determined the \gaia\ DR2 zero-point offset to be $52.3\pm2.0\,\uas$ when fitting a constant offset model. When allowing the \gaia\ offset to depend on $G$, $G_{BP}-G_{RP}$, and \teff, the result is shown in \eqnname~\eqref{eq:offset3} and \figurename~\ref{figure:offset}; the mean of this model is $\approx 50\,\uas$ with reasonable dependencies of the offset on $G$, $G_{BP}-G_{RP}$, and \teff\ that explain the differences between various zero-point determinations in the literature using different types of stars. Our zero-point determinations apply directly to all 139,847,389 stars in \gaia\ DR2 within the color--magnitude range covered by APOGEE. We release the catalog of spectro-photometric distances for APOGEE DR14 stars, with the data model given in \tablename~\ref{table:data_model} (see the Introduction for the data-file link). Moreover, all trained NN models from this work are publicly available (see Appendix~\ref{append:A}), including code to determine the zero-point offset for any \gaia\ star.

We compared our spectro-photometric distances to a subset of highly accurate \gaia\ parallaxes and to several alternative spectro-photometric distance catalogs. Our distances show excellent agreement with highly precise \gaia\ parallaxes, with distances to APOGEE RC stars, and with the distances to luminous red giants from \citet{2018arXiv181009468H}. But our distances give precise distances for a much wider range of stellar types and distances than any other existing method. We used the distances for the entire APOGEE DR14 sample and employed chemical abundances from our previous work \citep{2019MNRAS.483.3255L} to analyze spatial trends in 15 elemental abundance ratios across the Milky Way and find interesting trends across the Galaxy. In particular, we demonstrate that a bar-like trend is clearly apparent in many of the abundance ratios, consistent with a $\approx5\,\mathrm{kpc}$ half-length of the Galactic bar.

This method can be applied to future \gaia\ data releases and to future spectroscopic surveys.  Improving the determination of the true complex multivariate \gaia\ zero-point offset will be challenging and likely require inputting a good, physical model for the more complex dependencies of the zero-point (like its spatial dependence). But the \gaia\ zero-point offset will likely decrease in magnitude in future \gaia\ data releases and even with the current zero-point model, our method demonstrates that obtaining $\lesssim 10\%$ spectro-photometric distances is achievable for a wide range of main-sequence and giant stellar types.

\begin{table*}
	\centering
	\caption{Data Model of \texttt{apogee\_dr14\_nn\_dist.fits} which contains 277,371 APOGEE DR14 stars. Most labels contain $-9999$ for unknown or bad data.}
	\begin{tabular}{lccl} 
		\hline
		Label & Units & Sources & Descriptions \\
		\hline
		\texttt{apogee\_id} & n/a & APOGEE DR14 & APOGEE ID \\
		\texttt{location\_id} & n/a & APOGEE DR14 & Location ID of APOGEE \\
		\texttt{ra\_apogee} & deg & APOGEE DR14 & Right ascension (J2000) \\
		\texttt{dec\_apogee} & deg & APOGEE DR14 & Declination (J2000) \\
		\texttt{ra} & deg & Gaia DR2 & Right ascension (ICRS) \\
		\texttt{dec} & deg & Gaia DR2  & Declination (ICRS) \\
		\texttt{pmra} & \mas/yr & Gaia DR2 & Proper Motion in RA \\
		\texttt{pmdec} & \mas/yr & Gaia DR2 & Proper Motion in Dec \\
		\texttt{pmra\_error} & \mas/yr & Gaia DR2 & Proper Motion Uncertainty in RA  \\
		\texttt{pmdec\_error} & \mas/yr & Gaia DR2 & Proper Motion Uncertainty in Dec \\
		\texttt{phot\_g\_mean\_mag} & mag & Gaia DR2 & G-band mean magnitude \\
		\texttt{bp\_rp} & mag & Gaia DR2 & $G_{BP} - G_{RP}$ colour \\
		\texttt{fakemag} & mag & This work & NN Pseudo-luminosity $L_\mathrm{fakemag}$ (see \eqnname~\eqref{eq:fakemag})\\
		\texttt{fakemag\_error} & mag & This work & NN Pseudo-luminosity $L_\mathrm{fakemag}$ Uncertainty (see \eqnname~\eqref{eq:fakemag})\\
		\texttt{nn\_parallax}$^1$ & \mas & This work & NN Parallax (see \eqnname~\eqref{eq:fakemag})\\
		\texttt{nn\_parallax\_model\_error} & \mas & This work & NN Parallax Model Uncertainty (see \eqnname~\eqref{eq:fakemag})\\
		\texttt{nn\_parallax\_error} & \mas & This work & NN Parallax Total Uncertainty (see \eqnname~\eqref{eq:fakemag})\\
		\texttt{dist}$^1$ & parsec & This work & NN Inverse Parallax\\
		\texttt{dist\_model\_error} & parsec & This work & NN Inverse Parallax Model Uncertainty\\
		\texttt{dist\_error} & parsec & This work & NN Inverse Parallax Total Uncertainty\\
		\texttt{weighted\_dist}$^2$ & parsec & This work and Gaia DR2 & Inverse Variance Weighted Combined Distance from NN and Gaia\\
		& & & (NN Model Uncertainty is adapted) \\
		\texttt{weighted\_dist\_error} & parsec & This work and Gaia DR2 & Uncertainty in the Weighted Distance\\
		& & & (NN Model Uncertainty is adapted) \\
		\hline
	\end{tabular}
	\flushleft $^1$ Missing values have $-9999$, due to missing spectroscopic data or photometry. \\\vspace{-0.1in}
    \flushleft $^2$ Calculated based on the NN distance and $\plx_{\mathrm{Gaia}} +52\,\uas$. When one of the distance sources is $-9999$, the other one is used directly. When both are $-9999$, the resulting \texttt{weighted\_dist} is $-9999$.\\	
	\label{table:data_model}
\end{table*}

\section*{Acknowledgements}

HL and JB received support from the Natural Sciences and
Engineering Research Council of Canada (NSERC; funding reference number RGPIN-2015-05235) and from an Ontario Early Researcher Award (ER16-12-061). JB also received partial support from an Alfred P. Sloan Fellowship.

We gratefully acknowledge the support of NVIDIA Corporation with the donation of a Titan Xp GPU used in this research.

Funding for the Sloan Digital Sky Survey IV has been
provided by the Alfred P. Sloan Foundation, the U.S. Department
of Energy Office of Science, and the Participating
Institutions. SDSS-IV acknowledges support and resources
from the Center for High-Performance Computing at the
University of Utah. The SDSS web site is www.sdss.org.

This work has made use of data from the European Space Agency (ESA) mission
{\it Gaia} (\url{https://www.cosmos.esa.int/gaia}), processed by the {\it Gaia}
Data Processing and Analysis Consortium (DPAC,
\url{https://www.cosmos.esa.int/web/gaia/dpac/consortium}). Funding for the DPAC
has been provided by national institutions, in particular the institutions
participating in the {\it Gaia} Multilateral Agreement.







\appendix
\onecolumn

\section{Example of using Neural Net to infer luminosity and convert to distance on arbitrary APOGEE spectra}\label{append:A}

Besides providing general tools for deep learning in astronomy in the \texttt{astroNN} package, we also share the actual networks trained and discussed in this paper in a separate \texttt{GitHub} repository associated with this paper. Here we give an example of how to use the \emph{spectro-luminosity model} network for determining stellar luminosity and convert to distance for a given APOGEE spectrum. Moreover, code to use the \emph{Offset Calibration Model} separately to get the calibrated offset is not included in \texttt{astroNN}, but it is provided in the Github repository for this work. First follow the following instructions:
 
\begin{enumerate}
  \item Install \texttt{astroNN} by following instructions from \url{https://astronn.readthedocs.io/en/v1.1.0/quick_start.html}
  \item Obtain the repository containing the code to reproduce all figures in this paper at \url{https://github.com/henrysky/astroNN_gaia_dr2_paper}
  \item Open a python terminal under the repository folder but outside the neural network model folder
  \item Copy and paste the following code to do inference with the neural net in this paper on the star 2M19060637+4717296.
\end{enumerate}

\begin{lstlisting}[language=Python, caption={Example of using Neural Net to infer distance on APOGEE spectra}]
from astropy.io import fits
from astroNN.apogee import visit_spectra, apogee_continuum
from astroNN.gaia import extinction_correction, fakemag_to_pc
from astroNN.models import load_folder

# arbitrary spectrum
f = fits.open(visit_spectra(dr=14, apogee='2M19060637+4717296'))
spectrum = f[1].data
spectrum_err = f[2].data
spectrum_bitmask = f[3].data

# using default continuum and bitmask values to continuum normalize
norm_spec, norm_spec_err = apogee_continuum(spectrum, spectrum_err,
bitmask=spectrum_bitmask, dr=14)

# load neural net, it is recommend to use model ends with _reduced
# for example, using astroNN_constant_model_reduced instead of astroNN_constant_model
neuralnet = load_folder('astroNN_constant_model_reduced')

# inference, if there are multiple visits, then you should use the globally
# weighted combined spectra (i.e. the second row)
pred, pred_err = neuralnet.test(norm_spec)

# correct for extinction
K = extinction_correction(f[0].header['K'], f[0].header['AKTARG'])

# convert prediction in fakemag to distance
pc, pc_error = fakemag_to_pc(pred[:, 0], K, pred_err['total'][:, 0])
print(f"Distance: {pc} +/- {pc_error}")
\end{lstlisting}

\bsp	
\label{lastpage}
\end{document}